\documentclass[11pt, a4paper, oneside, reqno]{amsart}
\renewcommand\thesection{\arabic{section}}
\renewcommand\thesubsection{\thesection.\arabic{subsection}}

\makeatletter
\def\@settitle{\begin{center}%
		\baselineskip14\p@\relax
		\normalfont\LARGE\scshape\bfseries
		\@title
	\end{center}%
}
\makeatother

\makeatletter

\def\section{\@startsection{section}{1}%
	\z@{.7\linespacing\@plus\linespacing}{.5\linespacing}%
	{\normalfont\large\bfseries\centering}}

\def\subsection{\@startsection{subsection}{2}%
	\z@{.5\linespacing\@plus.7\linespacing}{.5\linespacing}%
	{\normalfont\large\bfseries}}

\def\subsubsection{\@startsection{subsubsection}{3}%
	\z@{.5\linespacing\@plus.7\linespacing}{.5\linespacing}%
	{\normalfont\itshape}}

\usepackage[colorlinks	= true,
			raiselinks	= true,
			linkcolor	= MidnightBlue,
			citecolor	= Mahogany,
			urlcolor	= ForestGreen,
			pdfauthor	= {Peyman Mohajerin Esfahani},
			pdftitle	= {},
			pdfkeywords	= {},   
			pdfsubject	= {},
			plainpages	= false]{hyperref}
\usepackage[usenames, dvipsnames]{color}
\usepackage{dsfont,amssymb,amsmath,subfig, graphicx,fancyhdr,mdframed}
\usepackage{amsfonts,dsfont,mathtools, mathrsfs,amsthm,wrapfig} 
\usepackage[]{siunitx}
\usepackage{ragged2e}

\let\labelindent\relax
\usepackage{enumitem}

\allowdisplaybreaks
\usepackage{algorithm}
\usepackage[noend]{algpseudocode}

\allowdisplaybreaks
\date{\today}

\patchcmd\@setauthors
  {\MakeUppercase{\authors}}
  {\authors}
  {}{}

\newtheorem{thm}{Theorem}[section]
\newtheorem{prop}{Proposition}

\newtheorem{assum}{Assumption}
\newtheorem{prob}{Problem}
\newtheorem{pset}{Problem Setting}
\newtheorem{subprob}{Sub-Problem}

\newtheorem{defn}{Definition}

\theoremstyle{remark}
\newtheorem{rem}{Remark}

\theoremstyle{remark}

\theoremstyle{definition}

\newcommand{\note}[1]{\textcolor{red}{(#1)}}

\newcommand{\EZ}[1][]{e}
\newcommand{\fEZ}[1][]{e}

\usepackage{optidef} %% to give number to constrtains in sdp 

\begin{document}
	
%	\begin{frontmatter}
		
		\title{ Model Updating for Nonlinear Systems \\ with Stability Guarantees} % Title, preferably not more 

%		\thanks[footnoteinfo]{Corresponding author F.~Ghanipoor.}

		\author{Farhad Ghanipoor$^{1}$, Carlos Murguia$^{1}$, Peyman Mohajerin Esfahani$^{2}$, Nathan van de Wouw$^{1}$}
		%\thanks{The authors are with the Department of Mechanical Engineering, Eindhoven University of Technology,($\{${\tt C.J.v.d.Ploeg, E.Silvas, N.v.d.Wouw@tue.nl$\}$@tue.nl}), the Delft Center for Systems and Control, Delft University of Technology ({\tt P.MohajerinEsfahani@tudelft.nl}), the Department of Civil, Environmental and Geo-Engineering, University of Minnesota, U.S.A., ({\tt nvandewo@umn.edu}) and the Netherlands Organisation for Applied Scientific Research, Integrated Vehicle Safety Group, 5700 AT Helmond, The Netherlands ($\{${\tt Chris.vanderPloeg, Emilia.Silvas$\}$@tno.nl}).}
		
	\thanks{1. Department of Mechanical Engineering, Eindhoven University of Technology, The Netherlands}
	\thanks{2. Delft Center for Systems and Control, Delft University of Technology, The Netherlands}

%	\begin{keyword}                           % Five to ten keywords,  
%	unmodeled dynamics;
%	machine learning;
%	input-to-state stability;
%	set invariance;
%	ultra local model;
%	\end{keyword}                            % keyword list or with the 
		% help of the Automatica 
		% keyword wizard
		\maketitle
		
		\begin{abstract}                          % Abstract of not more than 200 words.
					To improve the predictive capacity of system models in the input-output sense, this paper presents a framework for model updating via learning of modeling uncertainties in locally (and thus also in globally) Lipschitz nonlinear systems. First, we introduce a method to extend an existing known model with an uncertainty model so that stability of the extended model is guaranteed in the sense of set invariance and input-to-state stability. To achieve this, we provide two tractable semi-definite programs. These programs allow obtaining optimal uncertainty model parameters for both locally and globally Lipschitz nonlinear models, given uncertainty and state trajectories. Subsequently, in order to extract this data from the available input-output trajectories, we introduce a filter that incorporates an approximated internal model of the uncertainty and asymptotically estimates uncertainty and state realizations. This filter is also synthesized using semi-definite programs with guaranteed robustness with respect to uncertainty model mismatches, disturbances, and noise. Numerical simulations for a large data-set of a roll plane model of a vehicle illustrate the effectiveness and practicality of the proposed methodology in improving model accuracy, while guaranteeing stability.
		\end{abstract}
	
%	\end{frontmatter}

	\section{Introduction} \label{sec:intro}
	
	%*** at end double check name of theorems in the plots *** 
	%*** check equations name you wanted to force some tags for some  *** 
	
	%. 

	Dynamical systems modeling has been a key problem in many engineering and scientific fields, such as biology, physics, chemistry, and transportation. When modeling dynamical systems, it is of key importance to use well-established principles of physics and known prior system properties (e.g., stability and set-invariance) \cite{manchester2021contraction}. However, for many complex practical systems, we only tend to have partial knowledge of the physics governing their dynamics \cite{abbasi2022physics}. Even in cases where accurate physics-based models are established, as, e.g., in robotics, there exist inevitable (both parametric and non-parametric) uncertainties that impact the model's predictive accuracy.
	
	For a class of nonlinear dynamical systems, this paper assumes a prior system model is available (however, results here are also applicable when no prior model is available). It focuses on learning models for uncertainties while guaranteeing the stability of extended models (prior models plus uncertainty representations), given available input-output data. This problem contrasts with black box modeling approaches, such as Neural Networks (NNs) or Gaussian Processes (GPs), because we incorporate prior relationships derived from first principles into both the modeling and learning framework. Moreover, this problem differs from typical grey-box identification problems because, in our case, a prior model with known parameters is available. Off-the-shelf grey-box system identification methods can cope with a subset of the problem under discussion, where no prior model is available. The problem considered here enables the characterization of hybrid system representations (referred here as the extended model) comprising both prior physics-based and uncertainty learned models. In what follows, a comparison of our approach with some related existing literature on hybrid modeling (based on both first-principles and data) is provided.
	
	\emph{Existing Literature:} Our approach is fundamentally different from existing Physics-Informed (PI) learning techniques where standard black box models (such as NNs or GPs) are trained constrained to satisfy physics-based relations \cite{yazdani2020systems, daneker2023systems, taneja2022feature}. Yazdani et al. in \cite{yazdani2020systems} use this technique to construct the so-called Physics Informed Neural Networks (PINNs). They constrain a known physics-based model during the training of a NN-based model (i.e., penalize loss function for the mismatch between the physics-based model and the NN as a soft constraint) and incorporate physics knowledge in the structure of the NN. Although this method provides a NN as a system model (as well as parameters for physics-based models), it does not give a closed-form expression for the uncertainty in known physics-based models (i.e., it does not account for modeling mismatches due to unmodeled physics). Furthermore, parameters of both physics-based models and NNs are learned simultaneously, which increases the computational burden.

	The approach proposed in this paper also differs from the so-called Sparse Identification of Nonlinear Dynamics (SINDy) scheme \cite{brunton2016discovering}. SINDy assumes full knowledge of system states and their time derivatives. Then, based on known physics-based models and known variables (i.e., system states and their derivatives) a library of functions is generated that can be incorporated in dynamical models to account for uncertainty. To select the active functions in models, sparse identification algorithms are exploited. This approach has demonstrated accurate performance in sparse model identification of complex nonlinear systems \cite{champion2019data, champion2020unified, loiseau2018constrained}. However, SINDy not only requires full-state measurement but also requires the derivative of states to be known. Although the state derivatives can be approximated numerically if the complete state is known, most numerical methods are noise sensitive. Furthermore, the requirement of full-state measurements is a strong assumption for most dynamical systems. In our work, we do not require measurements of the full-state and its time derivative. The proposed algorithms need input-output data only.  
	
	Our approach augments a known physics-based model by a black-box model used as a correction term, see, e.g., \cite{schneider2022hybrid, bradley2022perspectives} and references therein. Such generic approach is also taken by Quaghebeur et al. in \cite{quaghebeur2021incorporating}, who add an NN model to a known physics-based model with unknown parameters. This approach allows maintaining the basic structure of the model that comes from first principles, which improves interpretability. However, simulating the hybrid model at each iteration during the training process is necessary. This approach is evidently more computationally intensive compared to our proposed method, which eliminates the requirement of simulating the model in every iteration. Moreover, the main drawback of this method is that it assumes that the initial state of the dynamic system is known or at least it requires measuring all the states (full-state measurement) of the true dynamical system. This assumption is dropped in the proposed method considered here. 
	
	Another important advantage of the proposed method is guaranteeing the stability of the extended nonlinear model (i.e., the model consisting of the known physics-based model and the uncertainty model). The identification of stable models has (mainly) been widely studied in the context of discrete LTI systems \cite{lacy2003subspace, di2023simba}. For further results on uncertainty learning for LTI systems, refer to \cite{ghanipoor2023uncertainty}. However, the identification of nonlinear stable models is still under study, mainly focusing on identifying the complete dynamics in a black box fashion. For instance, kernel- or Koopman-based methods that enforce some form of model stability during the learning process are proposed in \cite{khosravi2020nonlinear, khosravi2023representer} for the autonomous case and for the non-autonomous case in \cite{shakib2023kernel}. Moreover, there are results that aim to enforce model stability in other types of models, such as recurrent equilibrium network models \cite{revay2021recurrent} and Lur’e-type models \cite{shakib2019fast, revay2023recurrent}. However, it is important to note that none of these methods can be directly applied to the specific problem we are addressing here, given the \emph{prior known model}.
	
	In this paper, we propose a framework for model updating via learning modeling uncertainties in (physics-based) models applicable to locally Lipschitz nonlinear systems. We first focus on learning uncertainty models, assuming that some realizations of input, estimated uncertainty, and estimated state are given. During uncertainty learning, we guarantee that trajectories of the extended model (i.e., known prior model plus uncertainty model) belong to a given invariant set for locally Lipschitz extended models, or ensure input-to-state stability for globally Lipschitz extended models. This is achieved by formulating the problem as a constrained supervised learning problem.
	
	One key challenge in this problem involves the introduction of stability constraints, which is tackled using Lyapunov-based tools. The stability criteria usually result in an optimization problem that is non-convex. We address this challenge by proposing two different approaches for both locally and globally Lipschitz models:
	\begin{enumerate}
		\item \textbf{Cost Modification:} The first approach involves a change of variables, which leads to the rewriting of cost function (Theorems \ref{theorem:learning_modified_cost_locally} and \ref{theorem:learning_modified_cost_globally} for locally and globally Lipschitz models, respectively). 
		
		\item  \textbf{Constraint Modification:} This approach introduces a sufficient (convex) condition to satisfy the stability constraint (Theorems \ref{theorem:learning_modified_constraint_locally} and \ref{theorem:learning_modified_constraint_globally} for locally and globally Lipschitz models, respectively). 
		 
	\end{enumerate}
For the sake of completeness and comparison, we also provide a sequential algorithm that alternates the use of some variables in the optimization problem to convexity the program. However, we show in the numerical section that (as it is known in related literature \cite{boyd1994linear}) that initializing this algorithms is challenging, which further strengthens the importance of the  results provided here. We referred to the above mentioned sequential algorithm as the method of Sequential Convex Programming (SCP). 

\begin{figure}[t!]
		\centering
		\smallskip
		\includegraphics[width=0.9\linewidth,keepaspectratio]{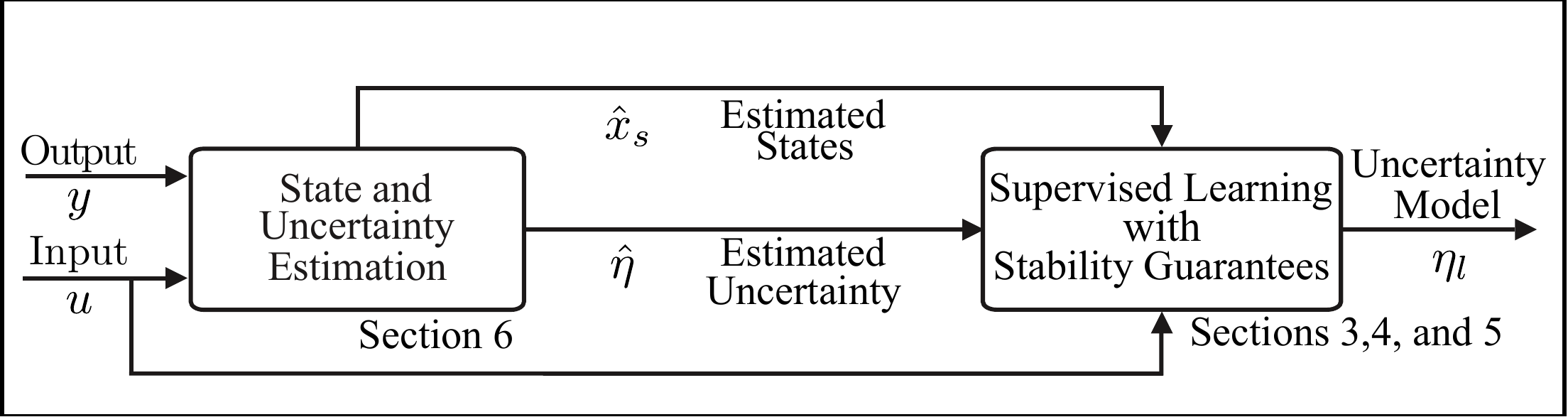}
		\caption{Overview of the MUNSyS methodology.}
		\label{fig:learning_overview}
	\end{figure}

%	However, the only initialization that yields a feasible solution for the SCP method is the solution of the proposed convex programs, which highlights the significance of these convex programs. 
	After addressing the challenge of non-convexity, the paper proceeds to discuss the practical implementation of the framework. It outlines a method for estimating uncertainty and state trajectories using input-output data and the known prior model (Proposition \ref{prop:optimal_estimator}). In this context, uncertainty is considered as an unknown input affecting the system dynamics and the estimation of uncertainty and state trajectories is achieved using robust state and unknown input observers \cite{ghanipoor2023robust}. For a schematic overview of the proposed methodology, Model Updating for Nonlinear Systems with Stability guarantees (MUNSyS), see Figure \ref{fig:learning_overview}. 
	
	In summary, the main contributions of this paper are as follows:
	\begin{enumerate} [label=(\alph*)]
		\item 	 \textbf{\emph{Practically Applicable Framework:}}  The proposed framework for learning modeling uncertainties in locally (and globally) Lipschitz nonlinear models only requires input-output data. Another important aspect for practical applicability is the lower computational cost compared to existing methods. This is achieved by eliminating the requirement to simulate the model in every iteration.
		\item \textbf{\emph{Stability Guarantee of Extended Models:}}  The proposed framework guarantees the following: 1. For locally Lipschitz extended models, trajectories of the extended model (i.e., known prior model plus uncertainty model) belong to a given invariant set; 2. For globally Lipschitz models, it ensures Input-to-State Stability (ISS).
		\item \textbf{\emph{Convex Approximations for Non-Convex Programs:}} Two distinct approaches for both locally and globally Lipschitz models, namely cost and constraint modification, are proposed to offer tractable convex approximations of formulated non-convex optimization problems for uncertainty model learning, while ensuring stability guarantees.
	\end{enumerate}

	This paper generalizes the preliminary results published in the conference paper \cite{ghanipoor2023uncertainty}. In comparison to \cite{ghanipoor2023uncertainty}, which is applicable to LTI systems only, we present results for a broader class of systems, considering both locally and globally Lipschitz nonlinear systems. The results in \cite{ghanipoor2023uncertainty} are a subclass of the results provided for globally Lipschitz nonlinear systems (see Theorem \ref{theorem:learning_modified_cost_globally}). 
	%Moreover, we assess the generalizability of learned models to unforeseen sets of data. Sequential CP
	
	\textbf{Notation:} 
	The set of nonnegative real numbers is represented by the symbol $\mathbb{R}^{+}$. The identity matrix of size $n \times n$ is denoted as $I_n$ or simply $I$ when the context specifies $n$. Similarly, matrices of dimensions $n \times m$ comprising only zeros are denoted as ${0}_{n \times m}$ or ${0}$ when the dimensions are clear. The first and second time-derivatives of a vector $x$ are denoted as $\dot{x}$ and $\ddot{x}$, respectively. For the $r^{th}$-order time-derivatives of vector $x$, the notation $x^{(r)}$ is employed. A positive definite matrix is symbolized by $X \succ 0$, and positive semi-definite matrices are indicated by $X \succeq 0$. Similarly, $X \prec 0$ is used for negative definite matrices, and $X \preceq 0$ for negative semi-definite matrices. The Hadamard (element-wise) power  of $n \in \mathbb{N}$ for a matrix $X$ is denoted by $X^{\circ n}$. The same notation is used for vectors. The $exp(x)$ shows the element-wise exponential function. The notation $(x_1, \ldots , x_n)$ signifies the column vector composed of elements $x_1, \ldots, x_n$, and this notation is also used when the components $x_i$ are vectors. Both Euclidean norm and the matrix norm induced by Euclidean norm are represented by the notation $||\cdot||$. The infinity-norm is denoted as $||\cdot||_\infty$. We employ the notation $\mathcal{L}_2(0, T)$ (or simply $\mathcal{L}_2$) to represent vector-valued functions $z: [0, T] \to \mathbb{R}^{k}$ satisfying $||z(t)||_{\mathcal{L}_2}^2 := \int_{0}^{T}||z(t)||^{2} dt < \infty$.
	%, and this notation is also used when $z$ is a vector-valued signal. 
	For a vector-valued signal $z$ defined for all $t \geq 0$, the $\mathcal{L}_\infty$ norm is denoted as $||z||_{\mathcal{L}_\infty} := \sup_{t \geq 0} ||z(t)||$. For a differentiable function $V: \mathbb{R}^{n} \to \mathbb{R}$, the row-vector of partial derivatives is denoted as $\frac{\partial V}{\partial x}$, and $\dot{V}(x)$ denotes the total derivative of $V(x)$ with respect to time (i.e., $\frac{\partial V}{\partial x} \frac{dx}{dt}$). The trace of a matrix $W$ is denoted as $tr(W)$. A continuous function $\bar \alpha:[0, a) \rightarrow[0, \infty)$ is said to belong to class $\mathcal{K}$ if it is strictly increasing and $\bar \alpha(0) = 0$. A continuous function $\bar \beta:[0, a) \times[0, \infty) \rightarrow[0, \infty)$ is said to belong to class $\mathcal{KL}$ if, for each fixed $s$, the mapping $\bar \beta(r,s)$ belongs to class  $\mathcal{K}$ with respect to $r$ and, for each fixed $r$, the mapping $\bar \beta(r,s)$ is decreasing with respect to $s$ and $\bar \beta(r, s) \rightarrow 0$ as $s \rightarrow \infty$. Time dependencies are often omitted for simplicity in notation.

	\section{Problem Formulation} \label{sec: problem formulation}
	Consider the nonlinear system
	\begin{equation} \label{eq:sys}
		\left\{\begin{aligned}
			\dot{x}_s &=A x_s+ B_u u+ S_g g(V_g x_s, u) + S_\eta \eta(V_\eta x_s, u) + B_\omega \omega,  \\
			y_s&= C x_s + D_\nu \nu,
		\end{aligned}\right. 
	\end{equation}
	where $x_s \in {\mathbb{R}^{n}}$, $y_s \in {\mathbb{R}^{{m}}}$, and $u \in {\mathbb{R}^{{l}}}$ are system state, measured output and known input vectors, respectively. Function $g: \mathbb{R}^{n_{v_g}} \times \mathbb{R}^{l} \to \mathbb{R}^{n_g}$ is a known nonlinear vector field, and function $\eta: \mathbb{R}^{n_{v_\eta}} \times \mathbb{R}^{l} \to \mathbb{R}^{n_\eta}$ denotes unknown modeling uncertainty. Signals $\omega: \mathbb{R}^{+}  \to \mathbb{R}^{n_\omega}$ and $\nu: \mathbb{R}^{+}  \to \mathbb{R}^{m_\nu}$ are unknown bounded disturbances; the former with unknown frequency range and the latter typically with high-frequency content (e.g., related to measurement noise). Known matrices $(A, B_u, S_g, V_g, S_\eta, V_\eta, B_\omega, C, D_\nu)$ are of appropriate dimensions, with ${n},{m}, l, n_{v_g}, n_g, n_{v_\eta}, n_{\eta}, n_{\omega}, m_\nu \in \mathbb{N}$. The matrices $S_g$ and $S_\eta$ specify the equations explicitly incorporating the nonlinearity $g$ and uncertainty $\eta$, while the matrices $V_g$ and $V_\eta$ identify the states influencing the nonlinearity and uncertainty, respectively. Moreover, without loss of generality, we assume that zero is an equilibrium point of the system for $u = 0$.
	
	In the following, we state the problem we aim to solve at a high abstraction level.
	
\begin{prob}\emph{\textbf{(Uncertainty Learning with Stability Guarantees)}} Consider system \eqref{eq:sys} with known input and output signals, $u$ and $y_s$. We aim to learn a data-based model for the uncertainty function $\eta(\cdot)$ so that the extended model composed of the known part of \eqref{eq:sys} and the learned uncertainty model are ``stable". The objective is to construct a more accurate system model (at least applicable to trajectories close to the training data set). 
	\label{prob:learning_high_level}
\end{prob}
	
	We first assume the availability of a labeled data-set containing input, estimated uncertainty, and estimated state realizations. A method to obtained this data from input-output trajectories is given in Section \ref{sec: uncertainty_state_estimation}. In what follows, we outline the problem settings for both locally and globally Lipschitz systems. 
	
	\subsection{Problem Settings} 
	
	We consider two model classes. For both model classes, the extended model (for the system in \eqref{eq:sys}) is of the form:
	\begin{equation} \label{eq:model}
		\begin{aligned}
			\dot{x} &= A x+ B_u u + S_g g(V_g x, u) + S_{\eta_l} \eta_l(V_\eta x, u), \\ 
			\eta_l(V_\eta x,u) &:= \Theta_l V_\eta x+B_l  u +  \Theta_n h(V_\eta x,u),
		\end{aligned}
	\end{equation}
	where $x \in {\mathbb{R}^{n}}$ is model state and function $\eta_l: \mathbb{R}^{n_{v_\eta}} \times \mathbb{R}^{l} \to \mathbb{R}^{n_{\eta_l}}$ is the uncertainty model that is parameterized by $\Theta_l, B_l$ and $\Theta_n$. Function $h: \mathbb{R}^{n_{v_\eta}} \times \mathbb{R}^{l} \to \mathbb{R}^{n_{h}}$ is a given nonlinear vector filed. This function contains the vector of basis functions that serve as candidates for the nonlinearities in the uncertainty model. Design matrices are collected as $\theta = (\Theta_l, B_l, \Theta_n)$ and have appropriate dimensions, $n_{\eta_l}, n_h \in \mathbb{N}$. Matrix $S_{\eta_l}$, similar to $S_\eta$ in \eqref{eq:sys}, indicates the explicit presence of the uncertainty model $\eta_l$ in the right-hand side and may differ from $S_{\eta}$. 
	
	\subsubsection{Cost Function} 
	Recall that, for now, we assume that estimated uncertainty and state realizations are available. Define the data vector corresponding to the $i$-th sample in time as
	\begin{equation*}
		\begin{aligned}
			d_i := {\left[\begin{array}{cccc}
					\hat{x}_i^\top V_\eta^\top & u_i^\top  & \hat{\eta}_i^\top  & h(V_\eta \hat{x}_i, u_i)^\top 
				\end{array}\right]^\top},
		\end{aligned}
	\end{equation*}
	where $\hat{x}_i, u_i,$ and $\hat{\eta}_i$ represent the state estimate, input, and uncertainty estimate, respectively. Vector $h(V_\eta \hat{x}_i, u_i)$ corresponds to the evaluated nonlinearity in \eqref{eq:model} at the given realizations (of state estimation and input). Given $N$ samples of the data vector defined above, define the data matrix $D$ as follows:
	\begin{equation} \label{eq:data_matrix}
		D := \sum_{i = 1}^{N} d_i d_i^\top.
	\end{equation}
	Further, consider the error vector between the uncertainty model and its (given) estimate as  
	\begin{equation} \label{eq:T}
		\begin{aligned}
			e_i := \eta_l(V_\eta \hat{x}_i, u_i)  - \hat{\eta}_i = T d_i, \quad \text{where}  \quad T := {\left[\begin{array}{cccc}
					\Theta_l  & B_l & -I & 	\Theta_n
				\end{array}\right]}.
		\end{aligned}
	\end{equation}
	Next, we introduce the following quadratic cost function to be minimized for the identification of $\Theta_l$, $B_l$, and $\Theta_n$:
	\begin{equation} \label{eq:quadratic_cost}
		\begin{aligned}
			J := \sum_{i = 1}^{N} e_i^\top e_i = \sum_{i = 1}^{N} d_i^\top T^\top T d_i.  
		\end{aligned}
	\end{equation}
	
	In the following sections, for two model classes, we provide stability constraints and formulate the constrained supervised learning of uncertainty models as optimization problems. 
	
	\subsubsection{Locally Lipschitz Model Class}  
		We first consider locally Lipschitz nonlinearities in \eqref{eq:model}, for both the basis functions $h(\cdot)$ and the known vector field $g(\cdot)$. Note that given bounded estimated state and input trajectories, we can always find ellipsoids that bound these trajectories. Hereafter, the system state and input trajectories are referred to as state and input sets, respectively. 
	
	Note that the set of all trajectories that can be generated by system in \eqref{eq:sys} is not known a priori. We only have finite data realizations of the estimated state in response to some input trajectories. Given this, we seek models \eqref{eq:model} that generate state trajectories ``close" to the available data set in response to inputs ``close" to the known input trajectories. To this end, we embed estimated state data and input trajectories in some known ellipsoidal sets, $\mathcal{E}_{sys}$ and $\mathcal{E}_{u}$, respectively. This embedding can be obtained efficiently, e.g., exploiting results in \cite[Sec. 2.2.4]{boyd1994linear}. 
	
\begin{figure}[t!]
		\centering
		\smallskip
		\includegraphics[width=0.6\linewidth,keepaspectratio]{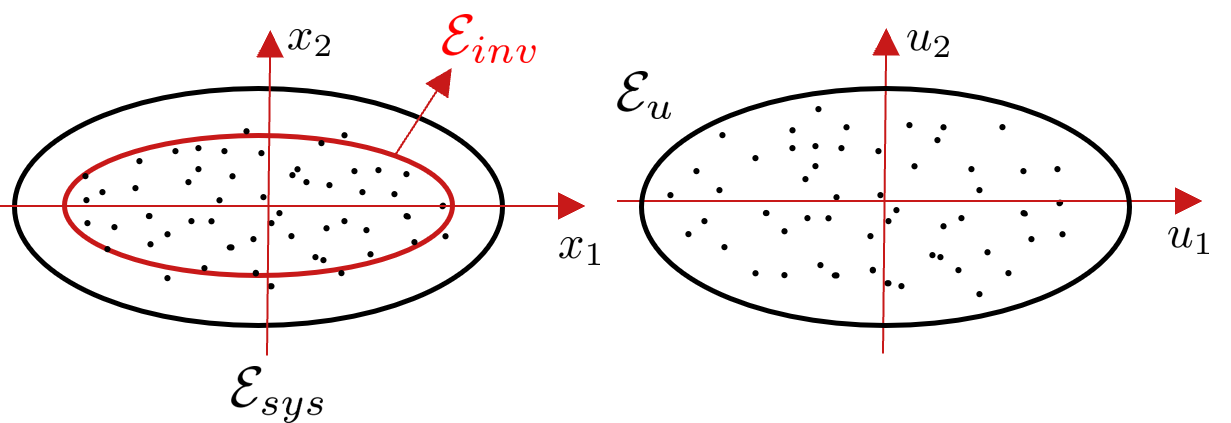}
		\caption{Illustration of different (ellipsoidal) sets for the system state $\mathcal{E}_{sys}$, input $\mathcal{E}_{u}$, and model state $\mathcal{E}_{inv}$.}
		\label{fig:sets}
	\end{figure}

	Having these ellipsoids $\mathcal{E}_{sys}$ and $\mathcal{E}_{u}$, to induce closeness of trajectories between system training data and model trajectories, we seek to enforce during learning that all trajectories generated by the model, in response to all input trajectories contained in $\mathcal{E}_{u}$, are contained in some known ellipsoidal set $\mathcal{E}_{inv}$. If we enforce the latter, and $\mathcal{E}_{inv} \subseteq \mathcal{E}_{sys}$, then we can guarantee that the model produces trajectories close to the training date set (close in the sense of set inclusion within $\mathcal{E}_{sys}$), see Figure~\ref{fig:sets}.
	
	Let the ellipsoidal set containing the state estimates training data be of the form:
	\begin{equation} \label{eq:system_set}
		\mathcal{E}_{sys} := \left\{x_s \mid x_s^{\top} F x_s \leq 1\right\},
	\end{equation}
	with $F \succeq 0$ of appropriate dimensions, and the input set of the form:
	\begin{equation} \label{eq:input_set}
		\mathcal{E}_{u} := \left\{u \mid u^{\top} U u \leq 1\right\},
	\end{equation}
	with $U \succeq 0$ of appropriate dimension.

	Because we seek to restrict model trajectories to the sets $\mathcal{E}_{u}$ and $\mathcal{E}_{sys}$, and consider models in \eqref{eq:model} with Lipschitz nonlinearities, we only require them to be Lipschitz within these sets. We formally state this in the following assumption. 
	
	\begin{assum}\emph{\textbf{(Locally Lipschitz Nonlinearities)}} The functions $g(V_g x, u)$ and $h(V_\eta x, u)$ in \eqref{eq:model} are locally Lipschitz in $\mathcal{E}_{u}$ and $\mathcal{E}_{sys}$, i.e., there exist known positive constants $l_{g_x}, l_{g_u}, l_{h_x}$ and $l_{h_u}$ satisfying
		\begin{equation} \label{eq:locally_lipschitz} 
			\begin{aligned}
				\|g(V_g x_1, u_1)-g(V_g x_2, u_2)\| &\leq  l_{g_x} \|V_g (x_1-x_2)\|+  l_{g_u}   \|u_1-u_2\|\,\\[2 mm]
				\|h(V_\eta x_1, u_1)-h(V_\eta x_2, u_2)\| &\leq  l_{h_x} \|V_\eta (x_1-x_2)\|\ 
				+  l_{h_u}   \|u_1-u_2\|\\
			\end{aligned}
		\end{equation}
		for all $x_1,x_2 \in  \mathcal{E}_{sys} \subset \mathbb{R}^{n}$ and all ${u_1, u_2} \in \mathcal{E}_{u} \subset \mathbb{R}^{l}$.
		\label{assum:locally_lipschitz}
	\end{assum}
	
In what follows, we formulate conditions to guarantee that model trajectories of \eqref{eq:model}, with locally Lipschitz nonlinearities, belong, in forward time, to an ellipsoidal invariant set:
\begin{equation} \label{eq:model_set}
	\mathcal{E}_{inv} := \left\{x \mid x^{\top} P x \leq 1\right\},
\end{equation}
guaranteeing $\mathcal{E}_{inv} \subseteq \mathcal{E}_{sys}$, for some positive definite matrix $P$. Conditions to ensure the latter can be formulated through Lyapunov-based stability tools and the S-procedure \cite[Sec. 2.6.3]{boyd1994linear}. For the sake of readability, these conditions are derived in Appendix \ref{ap:invariance_condtions}. There, it is shown that if the following conditions hold, the ellipsoid $\mathcal{E}_{inv}$ is a forward invariant set for \eqref{eq:model} and $\mathcal{E}_{inv} \subseteq \mathcal{E}_{sys}$:
	\begin{subequations} \label{eq:stability_locally}
		\begin{align}
			&\left[\begin{array}{ccc}
				\Delta+\beta P  & P (B_u+S_{\eta_l} B_l)  & 0 \\
				\star & (l_{g_u} + \bar l_{h_u}) I - \alpha U & 0 \\
				\star & \star & (\alpha - \beta) I
			\end{array}\right] \preceq 0, \label{eq:delta_locally_condition} \\
			& \left[\begin{array}{cc}
				\gamma     F -  P & 0 \\
				* & - \gamma +1
			\end{array}\right] \preceq 0, \label{eq:elipsoid_subset} \\
			&l_{h_x} \|  \Theta_n  \|  \leq  \bar l_{h_x},   \label{eq:theta_n_bound} 
		\end{align} 
		
		%\end{equation}
		where the involved components in the above equation are defined as
		\begin{equation} \label{eq:delta_globally} 
			\begin{aligned}
				\Delta &:= A^\top P+P A+  V_\eta ^\top \Theta_l^\top S_{\eta_l}^\top  P+ P S_{\eta_l} \Theta_l V_\eta + (l_{g_x} +l_{g_u}) P S_g S_g^\top P + (\bar l_{h_x} + \bar l_{h_u}) P S_{\eta_l} S_{\eta_l}^\top P \\
				&+l_{g_x} V_g^\top V_g + \bar l_{h_x} V_\eta^\top V_\eta,
			\end{aligned} 
		\end{equation}
		\begin{equation} \label{eq:l_h_bar_u} 
			\begin{aligned}
				\bar l_{h_u} &:= \bar l_{h_x} \frac{l_{h_u}}{l_{h_x}},
			\end{aligned} 
		\end{equation}
	\end{subequations}
	and $\bar l_{h_x}, \alpha, \beta, \gamma$ are adjustable parameters. In fact, $\bar{l}_{h_x}$ characterizes the size of a set within which $\Theta_n$ is enforced to reside. 
	
	Note that the condition in \eqref{eq:delta_locally_condition} is not convex in the invariant set shape matrix $P$ and uncertainty model parameters $\theta = (\Theta_l, B_l, \Theta_n)$, resulting in a non-convex optimization problem for uncertainty learning with invariance guarantees. Now, we can state the non-convex optimization problem for the locally Lipschitz model class, aiming to find a tractable convex solution later.
	
	\begin{pset}\emph{\textbf{(Locally Lipschitz Model Class)}} \label{prob:uncertainty_model_train_loaclly_lip}
		Consider a given data-set $D$ of input, estimated uncertainty, and state realizations and let Assumption~\ref{assum:locally_lipschitz} be satisfied. Further consider given ellipsoidal sets $\mathcal{E}_{sys}$ and $\mathcal{E}_{u}$, as introduced in \eqref{eq:system_set} and \eqref{eq:input_set}, respectively, containing the available input and state estimates data sets. Find the optimal parameters $\theta = (\Theta_l, B_l, \Theta_n)$ of the uncertainty model $\eta_l(\cdot)$ in \eqref{eq:model} (with locally Lipschitz nonlinearities) that minimizes the cost function $J$ in \eqref{eq:quadratic_cost} constrained to the ellipsoid $\mathcal{E}_{inv}$ in \eqref{eq:model_set} being a forward invariant set for the extended system model in \eqref{eq:model}. In other words, solve the non-convex optimization problem:
		\begin{mini}|s|        	% mini! = minimize 
			{\substack{P,\Theta_l, B_l,\Theta_n, \alpha, \gamma}}
			{&&&&J}   								% objective function and label
			{\label{eq:non_convex_optimization_loaclly_lip}}	
			{}   
			% optimization result
			\addConstraint{&&&&\eqref{eq:stability_locally} \quad \text{and} \quad P\succ 0.}{}   					  	% constraint   
		\end{mini}
	\end{pset}
	
In what follows, we formulate an analogue problem for the globally Lipschitz model class (that is assuming that the nonlinearities in \eqref{eq:model} are globally Lipschitz).
	
	\begin{rem} \emph{\textbf{(Comparison of Locally and Globally Lipschitz Model Classes)}} \label{rem:locally_globally_compar}
The main drawback of the globally Lipschitz model class is that it covers a smaller class of nonlinear systems compared to the locally Lipschitz class. However, we can enforce a stronger stability property (input-to-state stability \cite{sontag1995input}) for globally Lipschitz nonlinearities. Notably, for this model class, knowledge of the system state and input sets is no longer required. Furthermore, in the derivation of stability conditions for globally Lipschitz case, the need for a sufficient approximation (see S-procedure tools in Appendix \ref{ap:invariance_condtions}) is eliminated. This elimination relaxes the conservatism typically introduced by the sufficient condition. Moreover, the absence of this condition reduces the number of tuning parameters required for optimizing the globally Lipschitz class. The above arguments motivate to consider both model classes.
	\end{rem}
	
	\subsubsection{Globally Lipschitz Model Class}  
	For this model class, the nonlinearities in \eqref{eq:model} are assumed globally Lipschitz. 
	
	\begin{assum}\emph{\textbf{(Globally Lipschitz Nonlinearities)}} The functions $g(V_g x, u)$ and $h(V_\eta x, u)$ in \eqref{eq:model} are globally Lipschitz, i.e., there exist known positive constants $l_{g_x}, l_{g_u}, l_{h_x}$ and $l_{h_u}$ satisfying
		\begin{equation} \label{eq:globally_lipschitz} 
			\begin{aligned}
				\|g(V_g x_1, u_1)-g(V_g x_2, u_2)\| &\leq  l_{g_x} \|V_g (x_1-x_2)\| 
				+  l_{g_u}   \|u_1-u_2\|,\\[2 mm]
				\|h(V_\eta x_1, u_1)-h(V_\eta x_2, u_2)\| &\leq l_{h_x} \|V_\eta (x_1-x_2)\|
				+ l_{h_u}   \|u_1-u_2\|,\\
			\end{aligned}
		\end{equation}
		for all $x_1,x_2 \in {\mathbb{R}^{n}}$ and all ${u_1, u_2} \in {\mathbb{R}^{l}}$.
		\label{assum:globally_lipschitz}
	\end{assum}
	
In what follows, we formulate conditions to ensure Input-to-State Stability (ISS) of the extended model in \eqref{eq:model} satisfying Assumption \ref{assum:globally_lipschitz}. In the following definition, we introduce the notion of ISS for the extended model \cite{sontag1995input}. 
	
	\begin{defn} \emph{\textbf{(Input-to-State Stability)} The extended model \eqref{eq:model} is said to be ISS with respect to the input $u(t)$ if there exist a class $\mathcal{K} \mathcal{L}$ function $\bar \beta(\cdot)$ and a class $\mathcal{K}$ function $\bar \alpha(\cdot)$ such that for any initial state $x(t_0)$ and any bounded input $u(t)$, the solution $x(t)$ of \eqref{eq:model} exists for all finite $t \geq t_{0}$ and satisfies
			\begin{equation}\label{ISS_def}
				\|x(t)\| \leq \bar \beta\left(\left\|x\left(t_{0}\right)\right\|, t-t_{0}\right) + \bar \alpha( \sup _{t_0 \leq \tau \leq t}\|u (\tau)\| ).  
		\end{equation}}
	\end{defn}
	
	For readability, ISS conditions for the model in \eqref{eq:model} satisfying Assumption \ref{assum:globally_lipschitz} are derived in Appendix \ref{ap:iss_condtions}. It is shown that if the condition \eqref{eq:theta_n_bound} together with the following condition hold, 
	\begin{equation} \label{eq:delta_condition}
		\begin{aligned}
			&\Delta \prec 0,  \\
		\end{aligned} 
	\end{equation}
	where $\Delta$ is as defined in \eqref{eq:delta_globally}, then, the extended model in \eqref{eq:model} with globally Lipschitz nonlinearities is ISS with respect to the input $u(t)$.
	%and same as locally lightship model class, $\bar l_h$ is an adjustable parameter, which is selected for the minimal cost function in \eqref{eq:quadratic_cost}.  
	
	Now, we can state the non-convex optimization problem for the globally Lipschitz model class, aiming to find a tractable convex approximation later.
	
	\begin{pset}\emph{\textbf{(Globally Lipschitz Model Class)}} Consider a given data-set $D$ of input, estimated uncertainty, and state realizations and let Assumption \ref{assum:globally_lipschitz} be satisfied. Find the optimal parameters $\theta = (\Theta_l, B_l, \Theta_n)$ of the uncertainty model $\eta_l(\cdot)$ in \eqref{eq:model} (with globally Lipschitz nonlinearities) that minimizes the cost function $J$ in \eqref{eq:quadratic_cost}, such that the extended system model in \eqref{eq:model} is ISS with respect to input $u(t)$. In other words, solve the non-convex optimization problem:
		\begin{mini}|s|          	% mini! = minimize 
			{\substack{P,\Theta_l, B_l,\Theta_n}}                            % optimization variable
			{&&&&J}   								% objective function and label
			{\label{eq:non_convex_optimization_globally_lip}}             								% label for optimizatio problem
			{}                           								% optimization result
			\addConstraint{&&&&\eqref{eq:theta_n_bound}, \eqref{eq:delta_condition}, P\succ 0.}{}   					  	% constraint   
		\end{mini}
		%	with $\Delta$ as defined in \eqref{eq:delta_globally}, $l_{h_x}$ in \eqref{eq:globally_lipschitz}, and $\bar l_h$ in \eqref{eq:theta_n_bound}. 
		\label{prob:uncertainty_model_train_globally_lip}
	\end{pset}
	
	The challenge in Problem Settings \ref{prob:uncertainty_model_train_loaclly_lip} and \ref{prob:uncertainty_model_train_globally_lip} is that stability constraints are not convex, resulting in non-convex optimization problems. To tackle this challenge, we provide two approximate convex solutions for both problem settings using two distinct approaches: first, \emph{cost modification approach} and second, \emph{constraint modification approach}. Moreover, as an alternative solution, we provide a procedure to solve the original non-convex problems via sequential convex programming.
	
	\section{Cost Modification Approach} \label{sec: learning_sol_cost}
	We convexify the stability constraints by a change of variables and rewrite the cost function in \eqref{eq:quadratic_cost} in them. First, we provide the approximate solution for Problem Setting \ref{prob:uncertainty_model_train_loaclly_lip}, followed by the approximate solution for Problem Setting \ref{prob:uncertainty_model_train_globally_lip}.

	\subsection{Locally Lipschitz Model Class}
	The following theorem formalizes the associated convex optimization problem obtained via the cost modification approach as an approximation to the non-convex optimization problem in \eqref{eq:non_convex_optimization_loaclly_lip}. 
	
	\begin{thm}\textbf{\emph{(Stable Locally Lipschitz Model Learning with Modified Cost)}} 			\label{theorem:learning_modified_cost_locally}
		Consider system \eqref{eq:sys}, a given data-set $D$ of input, estimated uncertainty, and state realizations. In addition, consider given ellipsoidal sets $\mathcal{E}_{sys}$ and $\mathcal{E}_{u}$, as introduced in \eqref{eq:system_set} and \eqref{eq:input_set} with shape matrices $F \succeq 0$ and $U \succeq 0$, respectively. Further consider the extended system model in \eqref{eq:model}, under Assumption~\ref{assum:locally_lipschitz} with Lipschitz constants $l_{g_u}, l_{g_x}, l_{h_u}$ and $l_{h_x}$. Consider the following convex program:
		\begin{mini!}|s|[2]                 	% mini! = minimize 
			{\substack{P,S, R, Z, W, \alpha, \gamma}}                            % optimization variable
			{tr(W) \notag}   								% objective function and label
			{\label{eq:sdp_quadratic_cost_locally}}             								% label for optimizatio problem
			{}                           								% optimization result
			\addConstraint{\left[\begin{array}{ccccc}
					M_{11} & M_{12}  & 0 & M_{14} & M_{15}\\
					\star & M_{22} & 0 & 0 & 0\\
					\star & \star &  M_{33}& 0 & 0\\
					\star & \star & \star & -I & 0 \\
					\star & \star & \star & \star & -I
				\end{array}\right]}{\preceq 0 \label{eq:delta_convex_locally}}   					  	% constraint 1
			\addConstraint{\left[\begin{array}{cc}
					\bar l_{h_x}  I & l_{h_x} Z^\top\\
					* & \bar l_{h_x}  (2  \mu_1  P - \mu_1^2 I)
				\end{array}\right]	}{\succeq 0 \label{eq:theta_n_bound_convex}}
			\addConstraint{	\left[\begin{array}{cc}
					\gamma     F -  P & 0 \\
					* & - \gamma +1
				\end{array}\right]}{\preceq 0 \notag}
			\addConstraint{\left[\begin{array}{ccc}
					2 \mu_2 P & \tilde{T}\tilde{D}^\top & \mu_2 I \\
					* & I  & {0} \\
					* & * & W
				\end{array}\right]}{\succeq 0 \label{eq:cost_related_constraint_locally}}  						% constraint 2
				\addConstraint{P}{\succ 0, \quad \alpha, \gamma \geq 0\notag}   					  	% constraint 3
%			\addConstraint{\alpha, \gamma}{\geq 0 \notag}   					  	% constraint 3
		\end{mini!}
			with the involved matrices defined as follows:
		\begin{equation} \label{eq:delta_lmi_components} \tag{16d}
			\begin{aligned}
				M_{1} &:= A^\top P+P A+  V_\eta^\top S^\top + S V_\eta  +l_{g_x} V_g^\top V_g + \bar l_{h_x} V_\eta^\top V_\eta,  &M_{11} &:=  M_{1} + \beta P,\\
				M_{12} &:=	P B_u+R,
				&M_{22} &:= (l_{g_u} + \bar l_{h_u}) I - \alpha U, \\ 
				M_{14} &:=\sqrt{l_{g_x} +l_{g_u}}  P S_g, 
				&M_{15} &:=\sqrt{\bar l_{h_x} + \bar l_{h_u}} P, \\
				M_{33} &:=		 (\alpha - \beta) I, 
				&\tilde{T} &:= {\left[\begin{array}{cccc}
						S  & R & -P & Z
					\end{array}\right]},
			\end{aligned} 
		\end{equation}
		for given positive scalars $\bar l_{h_x}, \beta, \mu_1, \mu_2$, and $\bar l_{h_u}$ as defined in \eqref{eq:l_h_bar_u}, where $\tilde{D}$ is the Cholesky decomposition of the data matrix $D$ in \eqref{eq:data_matrix} (i.e., $D = \tilde{D}^\top \tilde{D}$), and the remaining matrices are the known parts of the system dynamics in \eqref{eq:sys}. Denote part of the optimizers of \eqref{eq:sdp_quadratic_cost_locally} as $P^\star$, $S^\star$, $R^\star,Z^\star$, and $W^\star$. Then, the following parameters of the extended model \eqref{eq:model}, $S_{\eta_l} = I, \Theta_l = \Theta_l^\star = P^{\star^{-1}} S^\star ,B_l = B_l^\star = P^{\star^{-1}} R^\star$ and $\Theta_n = \Theta_n^\star = P^{\star^{-1}} Z^\star$ guarantee that the ellipsoid $\mathcal{E}_{inv}$ in \eqref{eq:model_set} with $P = P^{\star}$ is a forward invariant set for the extended system model in \eqref{eq:model} and that $\mathcal{E}_{inv} \subseteq \mathcal{E}_{sys}$. In addition, it holds that the cost $J$ of \eqref{eq:non_convex_optimization_loaclly_lip} satisfies $J \leq tr(W^\star)$. As such, \eqref{eq:sdp_quadratic_cost_locally} represents an approximate convexified version of the problem in \eqref{eq:non_convex_optimization_loaclly_lip}.
	\end{thm}
	\emph{\textbf{Proof}:} 
	The proof can be found in Appendix \ref{ap:thm1_proof}.
	\hfill $\blacksquare$
	
	The scalar parameters $\bar l_{h_x}, \beta, \mu_1, \mu_2$, and $\bar l_{h_u}$  in Theorem~\ref{theorem:learning_modified_cost_locally} are tuned for the minimal feasible cost $tr(W)$ by a line search.
	%The following section provides results for the globally Lipschitz model class via the cost modification approach. 
	\subsection{Globally Lipschitz Model Class}
	The subsequent theorem formalizes the convex optimization problem using the cost modification approach, serving as an approximation to the non-convex optimization problem in \eqref{eq:non_convex_optimization_globally_lip}.
	\begin{thm}\textbf{\emph{(Stable Globally Lipschitz Model Learning with Modified Cost)}} 	\label{theorem:learning_modified_cost_globally}
		Consider system \eqref{eq:sys}, a given data-set $D$ of input, estimated uncertainty, and state realizations. In addition, consider the extended system model in \eqref{eq:model}, under Assumption~\ref{assum:globally_lipschitz} with Lipschitz constants $l_{g_u}, l_{g_x}, l_{h_u}$ and $l_{h_x}$. Consider the following convex program:
		\begin{mini!}|s|[2]        	% mini! = minimize 
			{\substack{P,S, R, Z, W}}                            % optimization variable
			{tr(W) \notag}   								% objective function and label
			{\label{eq:sdp_quadratic_cost_globally}}             								% label for optimizatio problem
			{}                           								% optimization result
			\addConstraint{\left[\begin{array}{ccc}
					M_{1}  & M_{14} & M_{15}\\
					\star & -I & 0 \\
					\star & \star & -I
				\end{array}\right]}{\prec 0 \label{eq:sdp_quadratic_cost_globally_1}}			  	% constraint 1
			\addConstraint{\left[\begin{array}{cc}
					\bar l_{h_x}  I & l_{h_x} Z^\top\\
					* & \bar l_{h_x}  (2  \mu_1  P - \mu_1^2 I)
				\end{array}\right]}{\succeq 0 \label{eq:sdp_quadratic_cost_globally_2}}
			\addConstraint{\left[\begin{array}{ccc}
					2 \mu_2 P & \tilde{T}\tilde{D}^\top & \mu_2 I \\
					* & I  & {0} \\
					* & * & W
				\end{array}\right]}{\succeq 0, \quad P \succ 0\notag}
%			\addConstraint{P}{ \notag}					  	% constraint 3
		\end{mini!}
		with $M_{1}, M_{14}, M_{15}, \tilde{T}$ as defined in \eqref{eq:delta_lmi_components}, for given positive scalars $\bar l_{h_x}, \mu_1,$ and $\mu_2$, where $\tilde{D}$ is the Cholesky decomposition of the data matrix $D$ in \eqref{eq:data_matrix}, and the remaining matrices are the known parts of the system dynamics in \eqref{eq:sys}. Denote part of the optimizers of \eqref{eq:sdp_quadratic_cost_globally} as $P^\star$, $S^\star$, $R^\star,Z^\star$, and $W^\star$. Then, the following parameters of the extended model \eqref{eq:model}, $S_{\eta_l} = I, \Theta_l = \Theta_l^\star = P^{\star^{-1}} S^\star ,B_l = B_l^\star = P^{\star^{-1}} R^\star$ and $\Theta_n = \Theta_n^\star = P^{\star^{-1}} Z^\star$ guarantee that the extended model in \eqref{eq:model} is ISS with respect to input $u(t)$. In addition, it holds that the cost $J$ of \eqref{eq:non_convex_optimization_globally_lip} satisfies $J \leq tr(W^\star)$. As such, \eqref{eq:sdp_quadratic_cost_globally} represents an approximate convexified version of the problem in \eqref{eq:non_convex_optimization_globally_lip}. As a special case, when the Lipschitz constants are zero (no nonlinearity) the conditions in \eqref{eq:sdp_quadratic_cost_globally_1} and \eqref{eq:sdp_quadratic_cost_globally_2} reduces to 
		\begin{equation} \label{eq:sdp_quadratic_cost_globally_linear} \tag{17c}
			A^\top P+P A+  V_\eta^\top S^\top + S V_\eta \prec 0.
		\end{equation}
	\end{thm}
	\emph{\textbf{Proof}:} 
	The proof follows the line of reasoning of the proof of Theorem~\ref{theorem:learning_modified_cost_locally} and it is omitted here due to space constraints. 
	\hfill $\blacksquare$
	
	\begin{rem} \emph{\textbf{(Surrogate Convex Optimizations with Modified Cost)}}
		We remark that the semi-definite programs in \eqref{eq:sdp_quadratic_cost_locally} and \eqref{eq:sdp_quadratic_cost_globally} are not equivalent to the non-convex optimization problems in \eqref{eq:non_convex_optimization_loaclly_lip} and \eqref{eq:non_convex_optimization_globally_lip}, respectively (i.e., these are convex approximations). This approximation is due to setting $S_{\eta_l} = I$ and using two sufficient conditions (two lower bounds) in the derivation of LMI conditions in each of the convex programs (see details at the end of the proof of Theorem \ref{theorem:learning_modified_cost_locally}). Although by letting $S_{\eta_l} = I$, we do not use the known structure of uncertainty, this makes the problem tractable. Note that here, we do not use knowledge of uncertainty structure. 
	\end{rem}

	Next, we follow a different approach to formulate alternative surrogate (approximate) convex optimization problems for non-convex optimization problems in Problem Settings \ref{prob:uncertainty_model_train_loaclly_lip} and \ref{prob:uncertainty_model_train_globally_lip}.
	
	\section{Constraint Modification Approach}  \label{sec: learning_sol_constraint}
	In this section, instead of changing the model-related variables in the stability constraints, we formulate sufficient conditions (upper bounds) for the stability constraints which are linear in all the optimization parameters. This convexifies the optimization problems in \eqref{eq:non_convex_optimization_loaclly_lip} and \eqref{eq:non_convex_optimization_globally_lip}. Similar to previous section, first, we provide the approximate solution for Problem Setting \ref{prob:uncertainty_model_train_loaclly_lip}, followed by the approximate solution for Problem Setting \ref{prob:uncertainty_model_train_globally_lip}.

	\subsection{Locally Lipschitz Model Class}
	
	The following theorem provides the convex approximation of the non-convex optimization problem in Problem Setting~\ref{prob:uncertainty_model_train_loaclly_lip} using the constraint modification approach.

	\begin{thm}\textbf{\emph{(Stable Locally Lipschitz Model Learning with Modified Constraint)}} 			\label{theorem:learning_modified_constraint_locally}
		Consider system \eqref{eq:sys}, a given data-set $D$ of input, estimated uncertainty, and state realizations. In addition, consider given ellipsoidal sets $\mathcal{E}_{sys}$ and $\mathcal{E}_{u}$, as introduced in \eqref{eq:system_set} and \eqref{eq:input_set} with characteristic matrices $F \succeq 0$ and $U \succeq 0$, respectively, and the extended system model in \eqref{eq:model}, under Assumption~\ref{assum:locally_lipschitz} with Lipschitz constants $l_{g_u}, l_{g_x}, l_{h_u}$ and $l_{h_x}$. Consider the following convex program:
		\begin{mini!}|s|[2]                 	% mini! = minimize 
			{\substack{Q,\Theta_l, B_l, \Theta_n, W, \alpha, \gamma}} % optimization variable
			{tr(W) \notag}   								% objective function and label
			{\label{eq:sdp_quadratic_constraint_locally}}             								% label for optimizatio problem
			{}                           								% optimization result
			\addConstraint{	\left[\begin{array}{cccccc}
					N_{11}  & N_{12} & 0 &N_{14} &  N_{15} & N_{16}\\
					\star   & N_{22} & 0 & 0 & 0 & 0 \\
					\star   &  \star & N_{33} & 0 & 0 & 0\\
					\star  & \star  &  \star  &  -2 \bar \gamma I & 0 & 0 \\
					\star  & \star  &  \star  & \star &  -I & 0 	\\
					\star  & \star  &  \star  &  \star &  \star & -I
				\end{array}\right]}{\preceq 0 \label{eq:delta_convex_locally_constraint_app}}   					  	% constraint 1
			\addConstraint{\left[\begin{array}{cc}
					\bar l_{h_x} I &  l_{h_x} \Theta_n^\top \\
					\star & \bar l_{h_x} I
				\end{array}\right] }{\succeq 0 \label{eq:theta_n_bound_convex_constraint_app}}
			\addConstraint{	\left[\begin{array}{cc}
					\gamma F -  2 \mu_3 I + \mu_3^2 Q & 0 \\
					* & - \gamma +1
				\end{array}\right]}{\preceq 0 \label{eq:elipsoild_subset_Q}}
			\addConstraint{tr(T D T^\top)}{\leq tr(W), \quad Q\succ 0, \quad \alpha, \gamma \geq 0  \label{eq:cost_constraint_modification}}   					  	% constraint 2
%			\addConstraint{ \notag}   					  	% constraint 3
%			\addConstraint{}{\notag}   					  	% constraint 3
		\end{mini!}
		with the involved matrices defined as follows:
		\begin{equation} \label{eq:delta_lmi_components_constraint_app} \tag{18e}
			\begin{aligned}
				N_{1} &:=  A Q +Q A^\top+ (l_{g_x} +l_{g_u}) S_g S_g^\top + (\bar l_{h_x}+ \bar l_{h_u}) S_{\eta_l} S_{\eta_l}^\top, &N_{11}& :=  N_{1} + \beta Q,\\
				N_{12} &:=	B_u+S_{\eta} B_l,
				&N_{14} &:= S_\eta \Theta_l V_\eta +\bar \gamma Q, \\
				N_{15} &:= \sqrt{l_{g_x}} Q V_g^\top, 
				&N_{16} &:= \sqrt{\bar l_{h_x}} Q V_\eta^\top, \\
				N_{22} &:= (l_{g_u} + \bar l_{h_u}) I - \alpha U, 
				&N_{33} &:=		 (\alpha - \beta) I,
			\end{aligned} 
		\end{equation}
		given positive scalars $\bar l_{h_x}, \beta, \mu_3, \bar \gamma$, and $\bar l_{h_u}$ as defined in \eqref{eq:l_h_bar_u}. The matrices $D, T$ are defined in \eqref{eq:data_matrix} and \eqref{eq:T}, respectively, and the remaining matrices are the known parts of the system dynamics in \eqref{eq:sys}. Denote part of the optimizers of \eqref{eq:sdp_quadratic_constraint_locally} as $Q^\star$, $\Theta_l^\star$, $B_l^\star$, $\Theta_n^\star$, and $W^\star$. Then, the following parameters of the extended model \eqref{eq:model}, $S_{\eta_l} = S_\eta, \Theta_l = \Theta_l^\star,B_l = B_l^\star$ and $\Theta_n = \Theta_n^\star$ guarantee that the ellipsoid $\mathcal{E}_{inv}$ as defined in \eqref{eq:model_set} with $P = Q^{{\star}^{-1}}$ is a forward invariant set for the extended system model in \eqref{eq:model}. In addition, it holds that the cost $J$ of \eqref{eq:non_convex_optimization_loaclly_lip} satisfies $J \leq tr(W^\star)$. As such \eqref{eq:sdp_quadratic_cost_locally} represents an approximate convexified version of the problem in \eqref{eq:non_convex_optimization_loaclly_lip}.
	\end{thm}
	\emph{\textbf{Proof}:} 
	The proof can be found in Appendix \ref{ap:thm3_proof}.
	\hfill $\blacksquare$

	\subsection{Globally Lipschitz Model Class}
	
	The subsequent theorem presents a formalization of the convex optimization problem derived through the constraint modification approach, serving as an approximation to the non-convex optimization problem in \eqref{eq:non_convex_optimization_globally_lip}.
	
	\begin{thm}\textbf{\emph{(Stable Globally Lipschitz Model Learning with Modified Constraint)}} 		\label{theorem:learning_modified_constraint_globally}
		Consider system \eqref{eq:sys}, a given data-set $D$ of input, estimated uncertainty, and state realizations. In addition, consider the extended system model in \eqref{eq:model}, under Assumption~\ref{assum:globally_lipschitz} with Lipschitz constants $l_{g_u}, l_{g_x}, l_{h_u}$ and $l_{h_x}$. Consider the following convex program:
		\begin{mini!}|s|[2]               	% mini! = minimize 
			{\substack{Q, \Theta_l, B_l, \Theta_n, W}}                            % optimization variable
			{&&&&tr(W) \notag}   								% objective function and label
			{\label{eq:sdp_quadratic_constraint_globally}}             								% label for optimizatio problem
			{}                           								% optimization result
			\addConstraint{&&&&\left[\begin{array}{cccc}
					N_{1}  & N_{14} & N_{15} & N_{16}\\
					\star & -2 \bar \gamma I & 0  & 0\\
					\star & \star & -I & 0 \\
					\star & \star &  \star & -I
				\end{array}\right]\prec 0}{\label{eq:sdp_quadratic_constraint_globally_1}}   					  	% constraint 1
			\addConstraint{&&&&\left[\begin{array}{cc}
					\bar l_{h_x} I &  l_{h_x} \Theta_n^\top \\
					\star & \bar l_{h_x} I
				\end{array}\right]\succeq 0}{\label{eq:sdp_quadratic_constraint_globally_2}}
			\addConstraint{&&&&tr(T D T^\top)}\leq tr(W),\quad Q\succ 0{\notag}  	
%			\addConstraint{&&&&{\notag}   					  	% constraint 3
		\end{mini!} 
		with $N_{1}, N_{14}, N_{15}, N_{16}$ as defined in \eqref{eq:delta_lmi_components_constraint_app}, given positive scalars $\bar l_{h_x}$ and $\bar \gamma$. The matrices $D, T$ are defined in \eqref{eq:data_matrix} and \eqref{eq:T}, respectively, and the remaining matrices are the known parts of the system dynamics in \eqref{eq:sys}. Denote part of the optimizers of \eqref{eq:sdp_quadratic_constraint_globally} as $Q^\star$, $\Theta_l^\star$, $B_l^\star$, $\Theta_n^\star$, and $W^\star$. Then, the following parameters of the extended model \eqref{eq:model}, $S_{\eta_l} = S_\eta, \Theta_l = \Theta_l^\star,B_l = B_l^\star$ and $\Theta_n = \Theta_n^\star$ guarantee that the extended model in \eqref{eq:model} is ISS with respect to input $u(t)$. In addition, it holds that the cost $J$ of \eqref{eq:non_convex_optimization_globally_lip} satisfies $J \leq tr(W^\star)$. As such \eqref{eq:sdp_quadratic_constraint_globally} represents an approximate convexified problem of the problem in \eqref{eq:non_convex_optimization_globally_lip}. As a special case, when the Lipschitz constants are zero (no nonlinearity) the conditions in \eqref{eq:sdp_quadratic_constraint_globally_1} and \eqref{eq:sdp_quadratic_constraint_globally_2} reduces to 
		\begin{equation} \label{eq:sdp_quadratic_constraint_globally_linear}
			\left[\begin{array}{cccc} \tag{19c}
				A Q +Q A^\top   & S_\eta \Theta_l +\bar \gamma Q\\
				\star & -2 \bar \gamma I\\
			\end{array}\right] \prec 0.
		\end{equation}
	\end{thm}
	\emph{\textbf{Proof}:} 
	The proof follows the line of reasoning of the proof of Theorem~\ref{theorem:learning_modified_constraint_locally} and is omitted for the sake of brevity. 
	\hfill $\blacksquare$
	
	%			add a remark here for scalars and refer after theorem 1 to here say what you said above for minimal cost and also say that tuning is easy.... so many numbers works... in most of cases small $\beta$ $10^{-3}$ and 1 for others.
	
	\begin{rem} \emph{\textbf{(Surrogate Convex Optimizations with Modified Constraint)}} \label{rem:constraint_app}
		The semi-definite programs in \eqref{eq:sdp_quadratic_constraint_locally} and \eqref{eq:sdp_quadratic_constraint_globally} are convex approximations of the non-convex optimization problems in \eqref{eq:non_convex_optimization_loaclly_lip} and \eqref{eq:non_convex_optimization_globally_lip}, respectively. Note that, a disadvantage of the constraint modification approach compared to the cost modification approach is that to ensure the feasibility of the semi-definite problems in \eqref{eq:sdp_quadratic_constraint_locally} and \eqref{eq:sdp_quadratic_constraint_globally}, the known $A$ matrix of the system in \eqref{eq:sys} has to be Hurwitz. Moreover, in deriving a convex approximation for the stability constraint, a non-dissipative term is substituted by a term that can potentially be dissipative (using Young's inequality). Thus, this solution may be more conservative compared to the cost modification approach (see for example, results in Section \ref{sec: sim results}).  On the other hand, unlike the cost modification approach, Theorems \ref{theorem:learning_modified_constraint_locally} and \ref{theorem:learning_modified_constraint_globally} use the knowledge of uncertainty structure by setting $S_{\eta_l} = S_\eta$, which is potentially beneficial. 
	\end{rem}

	In the following section, an alternative solution for solving the non-convex problems in Problem Settings \ref{prob:uncertainty_model_train_loaclly_lip} and \ref{prob:uncertainty_model_train_globally_lip} is provided using a procedure based on the technique of sequential convex programming.
	
	\section{Sequential Convex Program Approach}  \label{sec: scp}
	
	Another approach that can provide solution for the non-convex optimization problems in \eqref{eq:non_convex_optimization_loaclly_lip} and \eqref{eq:non_convex_optimization_globally_lip} is Sequential Convex Programming (SCP). This method typically involves fixing some parameters and solving a series of convex sub-problems iteratively \cite{boyd1994linear}. For the given non-convex optimization problems, we can fix Lyapunov function parameter $P$ or uncertainty model parameters $\theta = (\Theta_l, B_l, \Theta_n)$. The former transforms the non-convex optimization problem(s) in \eqref{eq:non_convex_optimization_loaclly_lip} (or in \eqref{eq:non_convex_optimization_globally_lip}) to feasibility problem(s) and the latter to convex optimization sub-problem(s). Then, this transformed problems can be solved iteratively until the convergence of the cost(s) of the convex optimization sub-problem(s). 
	
	Next, an SCP algorithm is presented as a sub-optimal solution for the non-convex optimization problem(s) in \eqref{eq:non_convex_optimization_loaclly_lip} (or in \eqref{eq:non_convex_optimization_globally_lip}). Steps 1 and 2 in this algorithm can be swapped to create a dual algorithm. In the loop below, initialization for $\theta$ is necessary, whereas for the dual loop, initialization for $P$ is required. SCP algorithm consisting of three steps:
	\begin{enumerate}[label=Step \arabic*:, leftmargin=*]
		\item Fix $\theta$ and find a feasible solution for the transformed feasibility problems in \eqref{eq:non_convex_optimization_loaclly_lip} (or \eqref{eq:non_convex_optimization_globally_lip});
		\item Fix $P$ as the feasible solution from Step 2 and solve the transformed sub-convex problem(s) in \eqref{eq:non_convex_optimization_loaclly_lip} (or \eqref{eq:non_convex_optimization_globally_lip}). 
		\item Redo Steps 1 and 2 until the cost of the transformed sub-convex problem(s) in Step 2 converge. Denote part of the optimizers of the transformed sub-convex problem(s) with $\theta = \theta^{\star}$ that provides the uncertainty model parameters. These uncertainty parameters guarantee stability of the extended model in \eqref{eq:model}. 
	\end{enumerate}
	
	%Note that among all possible initializations of the SCP, we have proposed the one above. The reason is, under the assumption that the known matrix $A$ of the system in \eqref{eq:sys} is Hurwitz, it can be easily seen that the existence of a feasible solution in Step 2 is guaranteed. Moreover, we remark that the convergence of the SCP is guaranteed since the cost function is uniformly lower bounded (here non-negative) and the sequence is non-increasing because in each step, you minimize the same function over some of the variables. It is clear that any monotone sequence with a uniform bound converges.
	
	\begin{rem} \emph{\textbf{(SCP and Convex Approximations Comparison)}} \label{rem:scp_convex_compar}
		The effectiveness of this SCP method is recognized to rely on the initialization \cite{boyd1994linear}. For example, in the simulation results in Section \ref{sec: sim results}, no other initialization was found except the solution of the convex approximation optimizations provided in Sections \ref{sec: learning_sol_cost} and \ref{sec: learning_sol_constraint}, highlighting the significance of the proposed convex programs. Moreover, depending on the optimization problem and initialization, the SCP method might require many iterations for convergence, making this method expensive in terms of computations. On the other hand, the drawbacks of the convex approximation programs are that they require some tuning and, due to conservatism in their derivation, their solutions might be more sub-optimal (i.e., have a larger cost) than the SCP method.
	\end{rem}
	
	In the results so far, we assumed that state and uncertainty realizations are available, which is, in practice typically not the case. In what follows, we present a solution for uncertainty and state estimation based on only input and output data.
	
	\section{Uncertainty and State Estimation} \label{sec: uncertainty_state_estimation}
	First, we formulate the uncertainty and state estimation problem before providing a solution for that problem. Consider system in \eqref{eq:sys} and the required assumptions below to ensure that the problem is well-posed as is common in the existing literature \cite{sontag2008input,han2019intermediate,Adaptive}. 
	
	\begin{assum}[Regularity]\label{assumption1}
		The following assumptions stand throughout this section: 
		\begin{itemize}
			
			\item 
			\emph{\textbf{State and Input Boundedness:}}  The state $x_s(t)$ and input $u(t)$ remain bounded over any finite time interval. 
			%	\label{assum:state_boundedness} 
			
			\item \emph{\textbf{$\mathcal{C}^r$ Uncertainty Vector:}} The uncertainty function $\eta(V_\eta x_s(t), u(t))$ in \eqref{eq:sys} is $r$ times differentiable with respect to time, i.e., the total time derivatives $\eta^{(1)}(V_\eta x_s(t),u(t))$,\linebreak$\eta^{(2)}(V_\eta x_s(t),u(t))$, ..., $\eta^{(r)}(V_\eta x_s(t),u(t))$ exist and are continuous.
			\label{assum:cr_assumption}

			\item \emph{\textbf{Bounded Disturbances:}} The disturbance vectors $\omega(t)$ and $\nu(t)$ in \eqref{eq:sys} are bounded on any finite time interval, and $\nu(t)$ is once differentiable in $t$, i.e., $\dot{\nu}(t)$ exists, and is continuous and bounded over any finite time interval.
			
		\end{itemize}
		
	\end{assum}
	
	The following filter is designed for uncertainty and state estimation:
	\begin{equation} 		\label{eq:filter}
		\left\{\begin{aligned} 
			\dot{z} =&f(z,u,y_s;\psi), \\
			\hat{\eta} =&\phi_1(z,y_s;\psi), \\
			\hat{x}_s =&\phi_2(z,y_s;\psi),
		\end{aligned}\right.
	\end{equation}
	where $z \in {\mathbb{R}^{n_z}}$ is the internal state of the filter with ${n_z} \in \mathbb{N}$. Functions $f: \mathbb{R}^{n_z} \times \mathbb{R}^{l} \times \mathbb{R}^{m} \to \mathbb{R}^{n_z}$, $\phi_{1}: \mathbb{R}^{n_z} \times \mathbb{R}^{m} \to \mathbb{R}^{n_\eta}$, and $\phi_{2}: \mathbb{R}^{n_z} \times \mathbb{R}^{m} \to \mathbb{R}^{n}$ characterize the filter structure, $\psi$ denotes design parameters. Define $\hat{x}_d := (\hat{\eta}, \hat{x}_s)$ (representing the estimate of both the uncertainty and the state) and its estimation error as 
	\begin{equation} \label{eq:ed}
		e_d :=\hat{x}_d - x_d,
	\end{equation}
	where ${x}_d := ({\eta}, {x}_s)$. Later, it will be shown that, for the uncertainty-state estimator design proposed in Section \ref{sec:observer}, the estimation error dynamics, explicated in Equation \eqref{eq:error_system} in Section \ref{sec:error_dynamics}, exhibits $(\omega,  \eta^{(r)},\nu, \dot{\nu})$ as a perturbation input. Having that in mind, we can state the problem we aim to solve at a high abstraction level.

	\begin{subprob}\emph{\textbf{(Uncertainty and State Estimation - Abstract Level)}} Consider the system \eqref{eq:sys} with known input and output signals, $u$ and $y_s$, and the uncertainty-state estimator filter \eqref{eq:filter}. For given $r$, design the filter parameters $\psi$ such that the following properties are guaranteed: \\
		\textbf{\emph{1) Stability:}} The estimation error dynamics is input-to-state stable with respect to the perturbation input $(\omega,  \eta^{(r)},\nu, \dot{\nu})$\emph{;}\\
		\emph{\textbf{2) Disturbance Attenuation:}} For $\nu = 0$, the $\mathcal{L}_2$-gain from $(\omega,  \eta^{(r)})$ to $e_d$ is bounded by some known $c_1 > 0$, for $t \geq 0$\emph{;}\\
		\emph{\textbf{3) Noise Rejection:}} For $ (\omega,  \eta^{(r)})= 0$, the $\mathcal{L}_2-\mathcal{L}_\infty$ induced gain from $(\nu, \dot{\nu})$ to $e_d$ is bounded by some known $c_2 > 0$, for $t \geq 0$\emph{.}
		\label{prob:uncertainty_state_estimation_high_level}
	\end{subprob}
	
	In what follows, we present a synthesis procedure to design the filter dynamics so that the three properties stated in Sub-Problem \ref{prob:uncertainty_state_estimation_high_level} are realized.
	
	\subsection{Ultra Local Uncertainty Representation}
	First some preliminaries are discussed, which are required to design the estimator filter \eqref{eq:filter}. Considering that the uncertainty $\eta(x_s(t), u(t))$ in \eqref{eq:sys} is an implicit function of time, for all $x_s(t)$ and $u(t)$, we can write an entry-wise $r$-th order Taylor time-polynomial approximation at time $t$ of $\eta$ as $\bar \eta = a_0 + a_1 t+ \dots + a_{r-1} t^{r-1}$ with coefficients $a_i \in {\mathbb{R}^{n_\eta}}, i= 0, \dots, r-1$. This model can be written in state-space form as
	\begin{equation} \label{eq:fault_model}
		\left\{\begin{aligned}
			\dot{\bar \zeta}_j &= \bar \zeta_{j+1},  \qquad 0 < j < r, \\
			\dot{\bar \zeta}_{r} &= 0,\\
			\bar \eta &= \bar \zeta_1,
		\end{aligned}\right.
	\end{equation}
	where $\bar \zeta_j \in {\mathbb{R}^{n_\eta}}$. Clearly, in the above model, we have $\bar \eta^{(r)} = {0}$, which might not be true for actual uncertainty signal $\eta$. Under Assumption \ref{assum:cr_assumption}, the actual internal state-space representation of $\eta$ is as follows:
	\begin{equation} \label{eq:fault_system}
		\left\{\begin{aligned}
			\dot{\zeta}_j &= \zeta_{j+1},  \qquad 0 < j < r, \\
			\dot{\zeta}_{r} &= \eta^{(r)},\\
			\eta &= \zeta_1,
		\end{aligned}\right.
	\end{equation}
	where $\zeta_j \in {\mathbb{R}^{n_\eta}}$. Clearly, the accuracy of the approximate model \eqref{eq:fault_model} increases as $\eta^{(r)}$ goes to zero (entry-wise). In the following, to design the uncertainty-state estimator we augment the system state, $x_s(t)$, with the states of the actual uncertainty internal state $\zeta_j(t), j \in\left\{1, \ldots, r\right\}$, and augment the system dynamics in \eqref{eq:sys} with \eqref{eq:fault_system}. We then design a filter (observer) for the augmented system to simultaneously estimate $x_s$ and $\zeta_j$ using model \eqref{eq:fault_model}. We remark that proper selection of the number of the uncertainty derivatives, $r$, added to the approximated model \eqref{eq:fault_model} (and \eqref{eq:fault_system}) is problem-dependent, see \cite{ghanipoor2022ultra} for discussion on selection of $r$.
	
	\subsection{Augmented Dynamics}
	Based on the internal representation of the uncertainty in \eqref{eq:fault_system}, define the augmented state $x_a:= (x_s, \zeta_1,\zeta_2,\ldots,  \zeta_{r})$, and rewrite the augmented dynamics using \eqref{eq:sys} and \eqref{eq:fault_system} as
	\begin{subequations} 		\label{eq:augmented}
		\begin{equation}		\label{eq:augmented_system}
			\begin{aligned}
				\left\{\begin{aligned}
					\dot{x}_{a} &=A_{a} x_{a}+B_{u_a} u+ S_{g_a} g(V_{g_a} x_{a}, u) +B_{\omega_a} \omega_{a},\\
					y_s &= C_{a} x_{a} +  D_{\nu} \nu,
				\end{aligned}\right.
			\end{aligned}
		\end{equation}
		where the involved matrices in the above equation are defined as
		\begin{equation} 		\label{eq:augmented_matrices}
			\begin{aligned}
				A_a &:= 
				\begin{bmatrix}
					A & S_\eta & {0} \\
					{0} & {0} & I_{d_n} \\
					{0} & {0} & {0} 
				\end{bmatrix}, \quad
				B_{u_a} :=	\left[\begin{array}{cc}
					B_u \\
					0
				\end{array}\right],  \quad 
				S_{g_a} := \left[\begin{array}{cc}
					S_g^\top &
					0
				\end{array}\right]^\top, \quad
				V_{g_a} := [\begin{array}[]{ll}V_g &0 \end{array}],\\
				B_{\omega_a} &:= \left[\begin{array}{cc}
					B_{\omega} & 0 \\
					0 & 0 \\
					0 &  I_{n_\eta}
				\end{array}\right], \quad
				\omega_{a} :=	\left[\begin{array}{cc}
					\omega \\
					\eta^{{(r)}} 
				\end{array}\right], \quad
				C_{a} := \left[\begin{array}{ll}	C	&0 \end{array}\right], \quad
				V_{\eta_a} := [\begin{array}[]{ll}V_\eta &0 \end{array}]
			\end{aligned}
		\end{equation}
		with $d_n := (r-1)n_{\eta}$.
	\end{subequations}

	\subsection{Uncertainty-State Estimator} \label{sec:observer}
	In this section, considering the uncertainty-state estimator general structure in \eqref{eq:filter}, inspired from observer-based approaches, we design $f(\cdot)$ and $\phi_i(\cdot), i=1,2$, as follows:
	\begin{subequations} \label{eq:observer}
		\begin{equation}
			\begin{aligned} 		\label{eq:observer_dynamics}
				f(z,u,y;\psi) &= N z+G u_a+L y_s +M S_{g_a} {g}(V_{g_a}\hat{x}_a+ H\color{black}(y_s - C_{a} \hat{x}_a),u), \\
				\phi_i (z,y;\psi) =& \bar{C}_i (z-E y_s), \\
			\end{aligned}
		\end{equation}
		with $\hat{x}_a = z-E y_s$, filter state $z \in {\mathbb{R}^{n_z}}$, $n_z = n+r n_{\eta}$, 
		\begin{equation*} 
			\bar{C}_1 := {\left[\begin{array}{ccc}
					{0} & I_{n_\eta} & {0} 
				\end{array}\right]}, \quad
			\bar{C}_2 := {\left[\begin{array}{cc}
					I_n & {0}
				\end{array}\right]},
		\end{equation*}
		and matrices $(N,G,L)$ defined as
		\begin{equation}		\label{eq:observer_matrices}
			\begin{aligned}
				N &:=M A_a-K C_a,
				&M&:=I+E C_a, \\[1mm]
				G&:=M B_a,
				&L&:=K(I+C_a E)-M A_a E.
			\end{aligned}
		\end{equation}
	\end{subequations}
	Matrices $E.K$ and $H$ are filter gains to be designed which can be collected as $\psi = (E,K,H)$. Note that according to \eqref{eq:observer_dynamics}, the part of the augmented state, $x_a$, that we use to reconstruct uncertainty and state signals is $\bar{C}_a x_{a}$ with
	\begin{equation} 	\label{eq:c_bar_a}
		\bar{C}_a := {\left[\begin{array}{cc}
				\bar{C}_1^\top &
				\bar{C}_2^\top 
			\end{array}\right]^\top}.
	\end{equation}
	In the following section, we analyze the estimator error dynamics.
	\subsection{Uncertainty-State Estimator Error Dynamics} \label{sec:error_dynamics}
	Consider the augmented state estimate $\hat{x}_a$ and let us define estimation error as 
	\begin{equation*}
		e:=\hat{x}_a-x_a=z-x_a-E y_s=z-M x_a-E D_\nu \nu.
	\end{equation*}
	Then, given the algebraic relations in \eqref{eq:observer_matrices}, the estimation error dynamics can be written as
	\begin{subequations}\label{eq:error_system}
		\begin{equation}	\label{eq:error_daymics_only}
			\left\{\begin{aligned}
				\dot{e} &= N e + M S_{g_a} \delta g - M B_{\omega_a} \omega_{a} + B_{\nu_a} \nu_a, \\
				e_d &= \bar{C}_a e,
			\end{aligned}\right.
		\end{equation}
		where the involved components in the above equation are defined as
		\begin{equation} \label{eq:error_dyamincs_related_matrices}
			\begin{aligned}
				\delta g &:= g(V_{g_a}\hat{x}_{a} + H (y_s - C_{a} \hat{x}_{a}), u)-g(V_{g_a} {x}_{a}, u), \\
				B_{\nu_a} &:= {\left[\begin{array}{cc}
						KD_\nu  & -E D_\nu 
					\end{array}\right]}, \qquad \nu_a := {\left[\begin{array}{cc}
						\nu^\top(t) & \dot{\nu}^\top(t)
					\end{array}\right]^\top}.
			\end{aligned}
		\end{equation}
	\end{subequations}
	
	Now, we can restate Problem \ref{prob:uncertainty_state_estimation_high_level} in a more formal way. 
	
	\begin{subprob}\emph{\textbf{(Uncertainty and State Estimation)}} Consider the system \eqref{eq:sys} with known input and output signals, $u$ and $y_s$. Furthermore, consider the internal uncertainty dynamics \eqref{eq:fault_system}, its Taylor approximation \eqref{eq:fault_model}, the augmented dynamics \eqref{eq:augmented}, the uncertainty-state estimator \eqref{eq:filter} with functions defined in \eqref{eq:observer}, and let Assumption \ref{assum:globally_lipschitz} be satisfied. Design the estimator gain matrices $\psi = (E,K, H)$ such that the following properties are guaranteed: \\
		\textbf{\emph{1) Stability:}} There exist a class $\mathcal{K} \mathcal{L}$ function $\bar \beta(\cdot)$ and a class $\mathcal{K}$ function $\bar \alpha(\cdot)$ such that for any initial estimation error $e(t_0)$ and any bounded input $\bar \omega_{a} :=(\omega , \eta^{(r)}, \nu, \dot{\nu})$, the solution $e(t)$ of \eqref{eq:error_system} exists for all $t \geq t_{0}$ and satisfies
		\begin{equation*}\label{ISS_def}
			\|e(t)\| \leq \bar \beta\left(\left\|e\left(t_{0}\right)\right\|, t-t_{0}\right) + \bar \alpha( \sup _{t_0 \leq \tau \leq t}\|\bar \omega_a (\tau)\| ) 
		\end{equation*}
		\emph{\textbf{2) Disturbance Attenuation:}} for $\nu=\dot{\nu}=0$, it holds that
		\begin{equation} \label{eq:J1}
			\tilde{J}_1(\psi) := \sup_{(\omega , \eta)} \frac{\|e_d\|_{\mathcal{L}_2}}{\|(\omega , \eta^{(r)})\|_{\mathcal{L}_2}}
		\end{equation}
		is bounded by some known $c_1 > 0$\emph{;}\\[1 mm]
		\emph{\textbf{3) Noise Rejection:}} for $\omega=\eta^{(r)}=0$, it holds that
		\begin{equation} \label{eq:J2}
			\tilde{J}_2(\psi) := \sup_{\nu} \frac{\|e_d\|_{\mathcal{L}_2}}{\|(\nu, \dot{\nu})\|_{\mathcal{L}_\infty}}
		\end{equation}
		is bounded by some known $c_2 > 0$\emph{.}
		\label{prob:uncertainty_state_estimation}
	\end{subprob}
	
	The essence of Problem \ref{prob:uncertainty_state_estimation} is to find an uncertainty-state estimator that, firstly,  ensures a bounded estimation error, $e(t)$, for any input (input to state boundedness); when $\nu = 0$ the $\mathcal{L}_2$-gain of the mapping from $(\omega , \eta^{(r)})$ to $e_d$ (the desired estimation error) is upper bounded by some $c_1>0$; and when $\omega=\eta^{(r)} = 0$, the $\mathcal{L}_2-\mathcal{L}_\infty$ induced gain from $(\nu, \dot{\nu})$ to $e_d$ is upper bounded by some $c_2>0$; and, for $\omega_{a} = {0}$ and $\nu_{a} = {0}$, $e(t)$ goes to zero asymptotically (internal stability).
	
	\subsection{Uncertainty-State Estimator Design}
	
	In the following proposition, we provide the solution of Problem \ref{prob:uncertainty_state_estimation} as a semi-definite problem, where we seek to minimize the $\mathcal{L}_2$-gain from $\omega_a$ to $e_d$ for an acceptable upper bound on the $\mathcal{L}_2-\mathcal{L}_\infty$ induced norm from $\nu_a$ to $e_d$ (there exist a trade-off between these two norms, see \cite{ghanipoor2023robust}). Moreover, we add the Input-to-State Stability (ISS) constraint with respect to filter error dynamics input $(\omega_a, \nu_a)$ to this program to enforce that stability of the resulting estimation filter. 
	
	\begin{prop}\emph{\textbf{(Estimator Design \cite{ghanipoor2023robust})}}
		Consider the system \eqref{eq:sys}, the augmented dynamics \eqref{eq:augmented}, the uncertainty-state estimator \eqref{eq:filter} with $f(\cdot)$ and $\phi(\cdot)$ as defined in \eqref{eq:observer}, and the corresponding estimation error dynamics \eqref{eq:error_system}, under Assumption~\ref{assum:globally_lipschitz} with Lipschitz constant $l_{g_x}$. Further, consider the following convex program:
		\begin{equation} 
			\begin{array}{cl}
				\min \limits_{\Pi, \bar R, \bar Q, H, \rho, \sigma}  & \rho \\
				\text{s.t.} & 						\vspace{1 mm}
				\left[\begin{array}{cc}
					X_{11} & X_{12} \\
					* & -I
				\end{array}\right] \prec 0,\\
				& \left[\begin{array}{ccc}
					L_{11} & -(\Pi + \bar R C_{a}) B_{\omega_a} & X_{12} \\
					* & -{\rho} a I  & 0 \\
					* & * & -I
				\end{array}\right]\preceq 0, \\[7mm]
				&\left[\begin{array}{cccc}
					X_{11} & H_{12} & 0 & X_{12}\\
					* & -b^2 I  & T_\nu^\top H^\top & 0\\
					* & * & -I & 0\\
					* & * & * & -I
				\end{array}\right]\preceq0, \\
				&\left[\begin{array}{cc}
					\Pi	 & \bar{C}_a^\top \\
					* & \sigma I
				\end{array}\right] \succeq 0, \quad \Pi \succ 0, \quad \sigma \leq \sigma_{max}\\
			\end{array}
			\label{eq:mimization}
		\end{equation}
		with the involved matrices defined as follows: 
		\begin{equation*}
			\begin{aligned}
				X_{11} &:= A_{a}^{\top}  \Pi +A_{a}^{\top} C_{a}^{\top} \bar R^{\top}-C_{a}^\top \bar Q^\top + \Pi A_{a} +\bar R C_{a} A_{a}  - \bar Q C_{a} + l_{g_x} (V_{g_a}^\top V_{g_a} -V_{g_a}^\top  H  C_{a} - C_{a}^\top  H^\top  V_{g_a}), \\
				X_{12} &:= {\left[\begin{array}{c}
						\sqrt{2 l_{g_x}}\big((\Pi+\bar R C_{a})S_{g_a}\big)^\top  \\  \sqrt{l_{g_x}}(H C_{a})^\top   
					\end{array}\right]^\top}, \\
				L_{11} &:=	X_{11}+ a \bar{C}_a^\top \bar{C}_a,\quad
				H_{12} := [\begin{array}[]{ll}  \bar Q D_\nu  & - \bar R D_\nu \end{array}], \quad   T_\nu  := [\begin{array}[]{ll}  D_\nu & 0 \end{array}], 
			\end{aligned}
		\end{equation*}
		given scalars $a, \sigma_{max} > 0$, $b$, matrix $\bar{C}_a$ as in \eqref{eq:c_bar_a}, and the remaining matrices in \eqref{eq:augmented_matrices} and \eqref{eq:sys}. Denote the optimizers as $\Pi^\star$, $\bar R^\star$, $\bar Q^\star$, $H^\star$, $\rho^\star$ and $\sigma^\star$. Then, the following parameters of \eqref{eq:observer}, $\psi= \psi^\star = \{E^\star = \Pi^{\star^{-1}} \bar R^\star ,K^\star = \Pi^{\star^{-1}} \bar Q^\star,H^\star\}$ guarantees the following:
		\begin{enumerate}
			\item The estimation error dynamics in \eqref{eq:error_system} is ISS with respect to perturbation input $(\omega_a, \nu_a)$;
			\item $\tilde{J}_1(\cdot)$ in \eqref{eq:J1} is upper bounded by $\sqrt{ \rho^\star}$, i.e., the $\mathcal{L}_2$-gain of \eqref{eq:error_system} with $\nu = 0$ from $\omega_a = (\omega , \eta^{(r)})$ to $e_d$ is upper bounded by $\sqrt{ \rho^\star}$;
			\item $\tilde{J}_2(\cdot)$ in \eqref{eq:J2} is upper bounded by $\sqrt{b \sigma^\star}$, i.e., the $\mathcal{L}_2-\mathcal{L}_\infty$ induced norm of \eqref{eq:error_system} with $\omega_a= 0$ from $\nu_a = (\nu, \dot{\nu})$ to $e_d$ is upper bounded by $\sqrt{b \sigma^\star}$.
		\end{enumerate}
		\label{prop:optimal_estimator}
	\end{prop}
	
	Note that the solution we have provided for uncertainty-state estimation is applicable to globally Lipschitz nonlinear systems (i.e., Assumption \ref{assum:globally_lipschitz} is satisfied). However, for locally Lipschitz nonlinear systems, a similar unknown input-state estimator proposed in \cite{ghanipoor2023linear} can be used.
	
	\section{Case Study}  \label{sec: sim results}
	
\begin{figure}[t!]
		\centering
		\smallskip
		\includegraphics[width=0.7\linewidth,keepaspectratio]{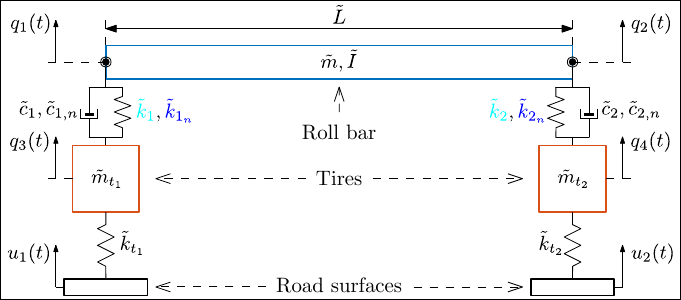}
		\caption{Roll plane system schematic.}
		\label{fig:rpm_schematic}
	\end{figure}

	In this section, we evaluate the proposed methods via a case study. We assess the obtained model accuracy for the considered class of nonlinear systems in the sense of output accuracy. The case study is a four-degree-of-freedom roll plane maneuver of a vehicle with nonlinear suspension, as considered in \cite{kessels2023real, blanchard2010polynomial}. See Figure \ref{fig:rpm_schematic} for a schematic of this system. The equations of motion and the system parameters are given in Appendix \ref{ap:rpm_app}. The nonlinearity of the system is rooted from nonlinear stiffness and damping between the roll bar and both tires. As available measured outputs, we have the relative displacement and relative velocity between the roll bar and both tires (i.e.,  $y_{s_1} = q_1 - q_3, y_{s_2} = q_2 - q_4, y_{s_3} = \dot{q}_1 - \dot{q}_3, y_{s_4} = \dot{q}_2 - \dot{q}_4$, see Figure \ref{fig:rpm_schematic}). The system inputs are the position of the tires ($u_1$ and $u_2$ in Figure \ref{fig:rpm_schematic}). The training and test data sets are generated by providing different system inputs, as described in Section \ref{sec:data_sets}.

	%The measurement noise is generated from a uniform distribution with an amplitude of five percent of the output signal for each sensor.  
	
	For the above-mentioned system, we have (unknown) parametric uncertainty in the linear stiffness between the roll bar and tires (i.e., uncertainty in the linear parameters $\tilde{k}_1, \tilde{k}_2$ shown in Figure \ref{fig:rpm_schematic}), and the nonlinear stiffnesses (i.e., $\tilde{k}_{1_n}, \tilde{k}_{2_n}$ shown in Figure \ref{fig:rpm_schematic}) are completely unknown. The problem is to find a data-based model for the mentioned uncertainties and guarantee stability of the extended model (i.e., known model together with uncertainty model). The proposed methods (three algorithms) presume that a (training) data-set of inputs, estimated uncertainties, and estimated state realizations is given (using uncertainty and state estimators discussed in Section \ref{sec: uncertainty_state_estimation}). The results of uncertainty and state estimation for one training data-set are provided in Appendix \ref{ap:estimation_result}.
	
	Following the proposed method for this problem, since the nonlinear stiffness is unknown, first we assume that we have basis functions as given by $h(\cdot)$ in \eqref{eq:model} that are representative of the unknown nonlinearity (the proposed method is evaluated for a range of basis functions in Section \ref{sec:basis_function_effect}). 
	%For this problem, given that the uncertainty is not a function of input signal, we have selected $B_l$ in uncertainty model \eqref{eq:model} as zero in the all of the results. Moreover, the basis function(s) are not a derived by input signal.  
	
	\subsection{Training and Test Data-Sets}  \label{sec:data_sets}
	
	The training and test data-sets are generated via different system inputs. System inputs comprise combinations of sinusoidals at different frequency ranges (to cover different range of inputs). This type of multi-sine signal is usual for system identification of the considered case study \cite[Sec. 3.3]{blanchard2010polynomial}. The system input $ u = (u_1, u_2)$ can be written as 
	\begin{equation*} \label{eq:input_sim}
		\begin{aligned}
			u_1 = \frac{\max {\tilde \alpha_i}}{\sum_{i=1}^{\tilde n} \tilde \alpha_i} \sum_{i=1}^{\tilde n} \tilde \alpha_i \sin \left(\tilde \omega_i t+\tilde \phi_i\right), \quad t \in[0, t_f]  \quad
			u_2 = 0,
		\end{aligned}
	\end{equation*} 
	 where the parameters 
	$$\left(\tilde \alpha_i\right)_{i=1}^{\tilde n}, \left(\tilde \omega_i\right)_{i=1}^{\tilde n}, \left(\tilde \phi_i\right)_{i=1}^{\tilde n},$$ 
	and ${\tilde n}$ are randomly drawn from uniform distributions in the intervals $[0.01,0.1]$ $m, [0.6 \pi,3 \pi]$ $rad/s , [0,0.94\pi]$ $rad,$ and $[2,10]$, respectively. The parameter $t_f$ indicates the final time of the simulation. 
	% his input can be interpreted as moving of a vehicle over multi sine surfaces (which can be characterized via $\tilde \alpha_i, \tilde \omega_i,\tilde \phi_i, \tilde n$ parameters) with different constant speeds (which can be characterized via $t_d$ parameter).
	Five training data sets and 1000 test data sets of inputs have been drawn from the aforementioned system input distributions.
		\begin{figure}[t!]
		\centering
		\smallskip
		\includegraphics[width=0.72\linewidth,keepaspectratio]{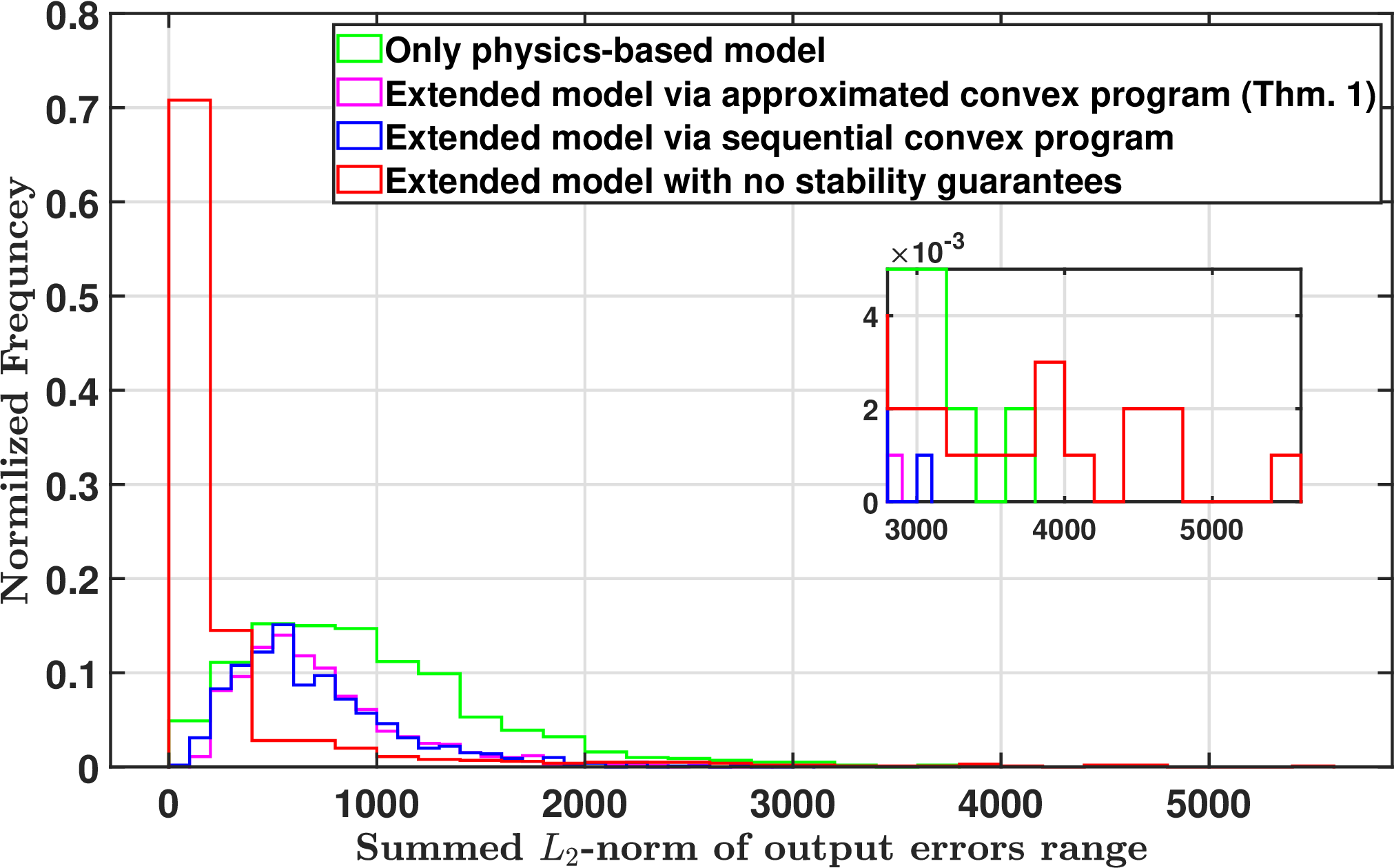}
		\caption{Histogram for comparison of system outputs and the outputs of different models with cubic basis function (i.e., $h = (V_\eta x)^{\circ 3}$) for the test data.}
		\label{fig:rpm_test}
	\end{figure}
	
	\subsection{Model Accuracy}  \label{sec:model_accuraty_rpm}
	
 Next, we compare the accuracy of the trained uncertainty models via the proposed methods. We compare the results against the true system model used to generate the data. We fit three extended models using the known physics-based model and the learned uncertainty models via the proposed methods described in Sections~\ref{sec: learning_sol_cost} and \ref{sec: scp}. Additionally, we examine a scenario where no stability guarantees for the extended model is considered (i.e., only by minimizing the cost function in \eqref{eq:quadratic_cost}). Thus, for the test data-set, we compare the system output with the outputs of three extended models and a model that is not extended and neglects uncertainties (i.e., only consist physics-based model).

 Given the large number of test data-sets (1000 test data-sets), this comparison is depicted via the histogram shown in Figure \ref{fig:rpm_test}. The horizontal axis is a metric for the output error computed via $\sum_{t=0}^{t_f = 20} \| y_s(t) - C x(t) \|$. The histogram presents the distribution of output errors for the test data by dividing it into intervals and displaying the frequency of occurrences within each interval. In the results shown in Figure \ref{fig:rpm_test}, the basis function representing the true system is a cubic function (i.e., $h = (V_\eta x)^{\circ 3} $ in \eqref{eq:model}). For results considering basis functions other than the cubic function, see Section \ref{sec:basis_function_effect}. From Figure \ref{fig:rpm_test}, one can conclude that both learning strategies (the results in blue and magenta) substantially enhance model quality compared to the model without a learned uncertainty model (green). Note that the SCP method is initialized with the solution of the approximated convex program. Other initializations, such as setting the parameters of the uncertainty model $\theta = (\Theta_l, B_l, \Theta_n)$ to zero or fixing the invariance set shape matrix $P$ as a constant multiplied by the identity matrix, did not yield a feasible solution. This further highlights the significance of the proposed convex approximation.
	
\begin{rem} \emph{\textbf{(The Price of Stability)}}
	Imposing the stability constraint turns out to deteriorate the expected output errors (compare the average of the red histogram with the blue and magenta histograms in Figure~\ref{fig:rpm_test}). The root cause of this can be due to the fact that the stability constraints make the resulting optimization program non-convex. Namely, we have to resort to a sub-optimal solution among all possible models considered (the magenta and blue histograms in Figure~\ref{fig:rpm_test}), whereas ignoring this condition allows us to stay in a convex setting where we can find the global optimal model (the red histogram). It is, however, worth noting that this better performance at the average level may come at the cost of some rare instances in the tail of the histogram (see the zoom window in Figure~\ref{fig:rpm_test}). In this light, one can see this average performance degradation as ``the price of stability" to ensure that any model we estimate meets our prior stability assumption.   
\end{rem}

	\subsection{Effect of Basis Functions}  \label{sec:basis_function_effect}
\begin{figure}[t!]
	\centering
	\smallskip
	\includegraphics[width=0.72\linewidth,keepaspectratio]{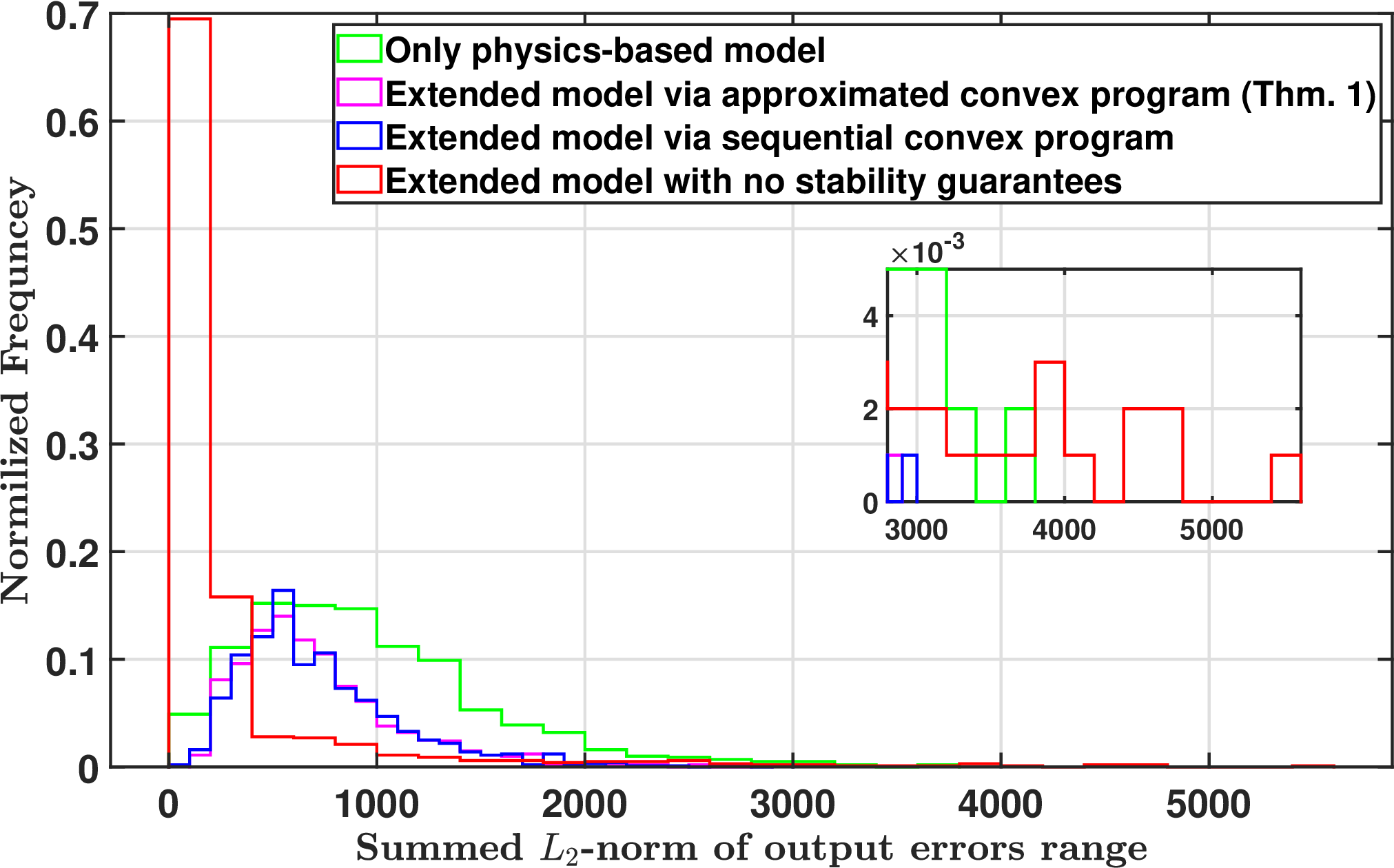}
	\caption{Histogram for comparison of system outputs and the outputs of different models with basis function $h = [(V_\eta x)^{\circ 2^\top} , (V_\eta x)^{\circ 3^\top}]^\top$ for the test data.}
	\label{fig:rpm_test_bf2}
\end{figure}

\begin{figure}[t!]
	\centering
	\smallskip
	\includegraphics[width=0.72\linewidth,keepaspectratio]{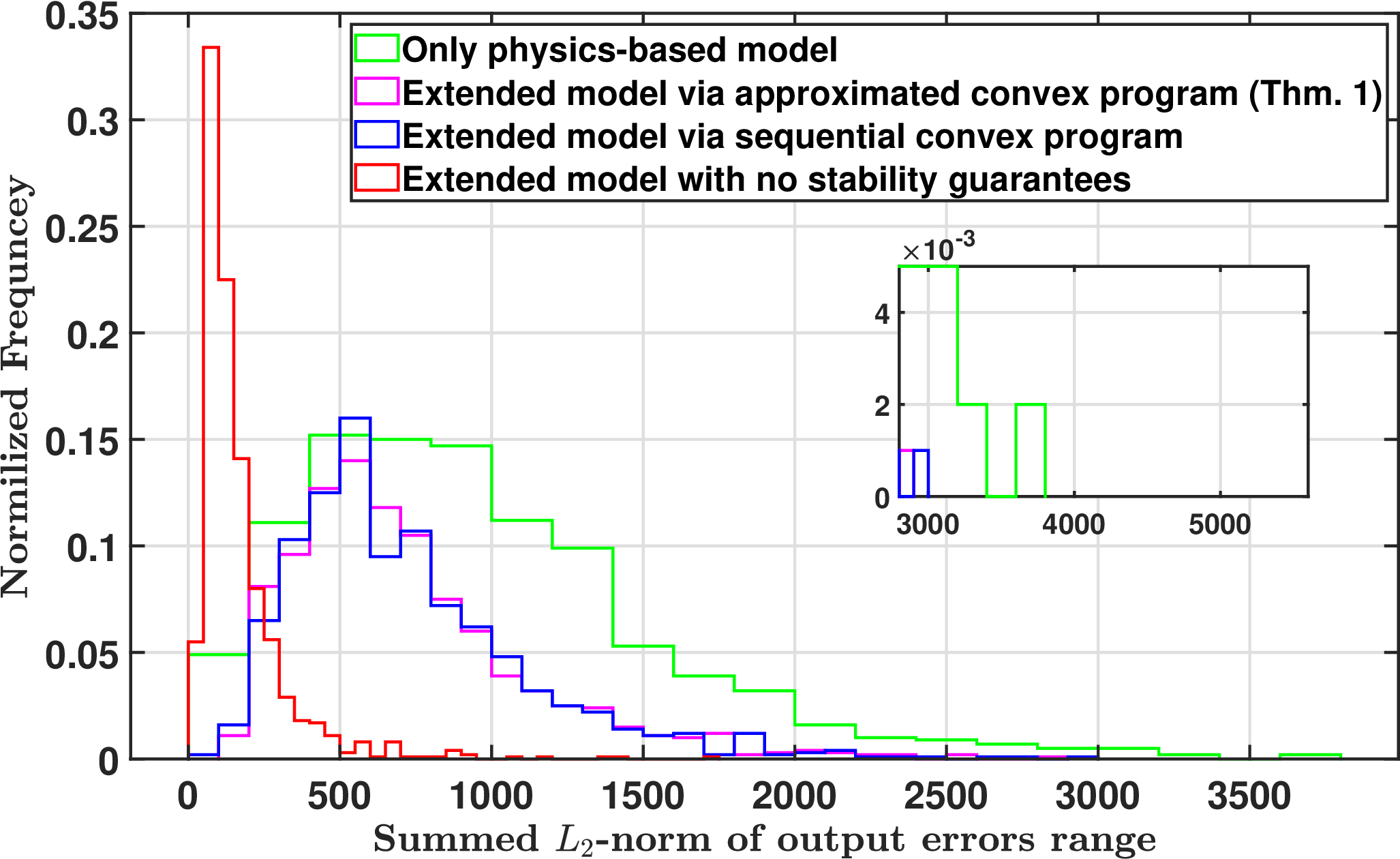}
	\caption{Histogram for comparison of system outputs and the outputs of different models with correct basis function $h = [(V_\eta x)^{\circ 2^\top}, exp(V_\eta x)^\top, (V_\eta x)^{\circ 3^\top}]^\top$ for the test data.}
	\label{fig:rpm_test_bf3}
\end{figure}

	Figures \ref{fig:rpm_test_bf2} and \ref{fig:rpm_test_bf3}, similar to Fig \ref{fig:rpm_test}, illustrate the histogram for the models output errors. Unlike Fig \ref{fig:rpm_test}, the basis functions (i.e., $h(\cdot)$ in \eqref{eq:model}) are not same as the true nonlinearity in the system. Now, the basis functions are $h = [(V_\eta x)^{\circ 2^\top} , (V_\eta x)^{\circ 3^\top}]^\top$ and $h = [(V_\eta x)^{\circ 2^\top}, exp(V_\eta x)^\top, (V_\eta x)^{\circ 3^\top}]^\top$ in Figures \ref{fig:rpm_test_bf2} and \ref{fig:rpm_test_bf3}, respectively. The results in these two figures indicate that the selection of the basis function does not significantly affect the accuracy of the extended model, especially for the results with stability guarantees. That is, having a model that uses the ``wrong" basis function is still better that using the model that does not account for uncertainty (at least in the considered use case).

		\section{Conclusion}\label{sec: conclusion}
		This paper has proposed a framework for model updating via learning modeling uncertainties in locally (and globally) Lipschitz nonlinear models. Moreover, we have ensured the stability of the extended model, which consists of a prior known model and the learned uncertainty model. The proposed framework includes two steps. First, we have assumed that uncertainty and state estimates are known. Under this assumption, we have introduced two semi-definite programs: Theorems \ref{theorem:learning_modified_cost_locally} and \ref{theorem:learning_modified_constraint_locally} for locally (and Theorems \ref{theorem:learning_modified_cost_globally} and \ref{theorem:learning_modified_constraint_globally} for globally) Lipschitz nonlinear models, to learn uncertainty models while ensuring the stability of the extended model (invariant set and ISS for locally and globally Lipschitz nonlinear models, respectively). Second, we have proposed a filter, designed using the semi-definite program outlined in Proposition \ref{prop:optimal_estimator}, to estimate uncertainty and state, based on the known prior model and input-output data. Simulations for a large data-set of a roll plane model of a vehicle have demonstrated the performance and potential of the proposed approach. 
		
	\appendix
\section*{Appendix}  
		\section{Invariant set conditions for model \eqref{eq:model}}   \label{ap:invariance_condtions}
		Let $W(x) = x^\top P x$ with $P \succ 0$ be a Lyapunov function candidate. We need to derive the derivative of the Lyapunov function candidate along the trajectories of the extended system model \eqref{eq:model}. Before that, first, we impose the condition $\mathcal{E}_{inv} \subseteq \mathcal{E}_{sys}$ on the model invariant set and the system state set as defined in \eqref{eq:model_set} and \eqref{eq:system_set}, respectively. By doing so, later, the Lipschitz property of the nonlinearities of extended system model \eqref{eq:model} can be used in deriving a tractable condition for $\dot{W}~\leq~0$. It is shown in \cite[Sec. 3.7.1]{boyd1994linear} that the condition $\mathcal{E}_{inv} \subseteq \mathcal{E}_{sys}$ can be written as \eqref{eq:elipsoid_subset} by applying $\mathcal{S}$-procedure. 
		
		From \eqref{eq:model} and the Lipschitz conditions for the known nonlinearity $g(\cdot)$ and the uncertainty model nonlinearity $h(\cdot)$ in \eqref{eq:locally_lipschitz} (together with that zero is an equilibrium point of system for $u=0$), it follows that 
		\begin{equation}		
			\begin{aligned} \dot{W}(x) = &x^\top \left(A^\top P+P A+  V_\eta ^\top \Theta_l^\top S_{\eta_l}^\top  P+ P S_{\eta_l} \Theta_l V_\eta \right) x+2 x^\top P ( S_g g( V_g x,u)+S_{\eta_l} \Theta_n h( V_\eta x,u)) \\ 
				& +2 x^\top P (B_u +S_{\eta_l} B_l)u\\
				\leq &x^\top \left(A^\top P+P A+  V_\eta ^\top \Theta_l^\top S_{\eta_l}^\top  P+ P S_{\eta_l} \Theta_l V_\eta \right) x+2 \|x^\top P S_g g( V_g x,u) \|\\
				&+ 2 \|x^\top P S_{\eta_l} \Theta_n h( V_\eta x,u)\| +2 x^\top P (B_u +S_{\eta_l} B_l)u\\
				\leq &x^\top \left(A^\top P+P A+  V_\eta ^\top \Theta_l^\top S_{\eta_l}^\top  P+ P S_{\eta_l} \Theta_l V_\eta \right) x +2 \|x^\top P S_g \| \| g( V_g x,u) \|\\
				&+ 2 \|x^\top P S_{\eta_l} \| \| \Theta_n h (V_\eta x,u)\| +2 x^\top P (B_u +S_{\eta_l} B_l)u\\
				\leq &x^\top \left(A^\top P+P A+  V_\eta ^\top \Theta_l^\top S_{\eta_l}^\top  P+ P S_{\eta_l} \Theta_l V_\eta \right) x +2 \|x^\top P S_g \|  (l_{g_x} \|V_g x\| + l_{g_u} \|u\|) \\
				&+ 2 \|x^\top P S_{\eta_l} \| \| \Theta_n \| (l_{h_x} \|V_\eta x\| + l_{h_u} \|u\|) +2 x^\top P (B_u +S_{\eta_l} B_l)u.
				\label{eq:lyapanov_ap_1}
			\end{aligned}
		\end{equation}
	
		Now, by imposing the condition on parameters of the nonlinear part of uncertainty model given in \eqref{eq:theta_n_bound} and then a specific form of the Cauchy–Schwarz inequality an upper bound for the right-hand-side of \eqref{eq:lyapanov_ap_1} can be found as follows:
		\begin{equation}		
			\begin{aligned}
				\dot{W}(x)\leq &x^\top \left(A^\top P+P A+  V_\eta ^\top \Theta_l^\top S_{\eta_l}^\top  P+ P S_{\eta_l} \Theta_l V_\eta \right) x+2 \|x^\top P S_g \|  (l_{g_x} \|V_g x\| + l_{g_u} \|u\|) \\
							&+ 2 \|x^\top P S_{\eta_l} \| (\bar l_{h_x} \|V_\eta x\| + \bar l_{h_u} \|u\|) +2 x^\top P (B_u +S_{\eta_l} B_l)u\\
			\leq &x^\top \left(A^\top P+P A+  V_\eta ^\top \Theta_l^\top S_{\eta_l}^\top  P+ P S_{\eta_l} \Theta_l V_\eta \right) x+ (l_{g_x} +l_{g_u}) \|x^\top P S_g \|^{2}+ l_{g_x}\|V_g x\|^{2} + l_{g_u} \|u\|^{2} \\
				& + (\bar l_{h_x} +\bar l_{h_u}) \|x^\top P S_{\eta_l} \|^{2}+ \bar l_{h_x} \|V_\eta x\|^{2} + \bar l_{h_u} \|u\|^{2} +2 x^\top P (B_u +S_{\eta_l} B_l)u\\
				= &x^\top\big(A^\top P+P A+  V_\eta ^\top \Theta_l^\top S_{\eta_l}^\top  P+ P S_{\eta_l} \Theta_l V_\eta + (l_{g_x} +l_{g_u}) P S_g S_g^\top P + (\bar l_{h_x} + \bar l_{h_u}) P S_{\eta_l} S_{\eta_l}^\top P \\
				&+l_{g_x} V_g^\top V_g + \bar l_{h_x} V_\eta^\top V_\eta \big) x + (l_{g_u} + \bar l_{h_u}) u^\top u  +2 x^\top P (B_u +S_{\eta_l} B_l)u  \\
				= & x^\top \Delta x + 2 x^\top P (B_u+S_{\eta_l} B_l) u +(l_{g_u} + \bar l_{h_u}) u^\top u
			\end{aligned}
			\label{eq:lyapanov_ap}
		\end{equation}
		
		with $\bar l_{h_u}$ as defined in \eqref{eq:l_h_bar_u} and $\Delta$ as defined in \eqref{eq:delta_globally}. Therefore, from \eqref{eq:lyapanov_ap}, it follows that 
		\begin{equation}
			\begin{aligned} \dot{W}(x) \leq& x^\top \Delta x + 2 x^\top P (B_u+S_{\eta_l} B_l) u +(l_{g_u} + \bar l_{h_u}) u^\top u.
			\end{aligned}
			\label{eq:lyapanov}
		\end{equation}
		
		If we can find a $P$ such that $\dot{W} \leq 0$ whenever $u \in \mathcal{E}_{u}$ (as defined in \eqref{eq:input_set}) and $W \geq 1$ along the trajectories of \eqref{eq:model}; Then, the ellipsoid $\mathcal{E}_{inv}$ (as defined in \eqref{eq:model_set}) is an invariant set of \eqref{eq:model} \cite{lin2022plug}. To this end, consider the vector $\left[x^\top u^\top 1 \right]^\top$. Using \eqref{eq:lyapanov_ap}, the condition $\dot{W} \leq 0$ can be stated as 
		\begin{equation} \label{eq:vdot_cond}
			E_1 = \left[\begin{array}{ccc}
				\Delta  & P (B_u+S_{\eta_l} B_l)  & 0 \\
				\star & (l_{g_u} + \bar l_{h_u}) I & 0 \\
				\star & \star & 0
			\end{array}\right] \preceq 0.
		\end{equation}
		Similarly, the condition \eqref{eq:input_set} can be written as 
		\begin{equation} \label{eq:u_cond}
			F_1 =\left[\begin{array}{ccc}
				0  & 0 & 0 \\
				\star & U & 0 \\
				\star & \star & -1
			\end{array}\right] \preceq 0.
		\end{equation}
		Finally, the condition $W(x) \geq 1$ can be restated as 
		\begin{equation} \label{eq:v_cond}
			G_1 =\left[\begin{array}{ccc}
				-P  & 0 & 0 \\
				\star & 0 & 0 \\
				\star & \star & 1
			\end{array}\right] \preceq 0.
		\end{equation}
		Therefore, for constructing an invariant set of trajectories, we require \eqref{eq:vdot_cond} ($\dot{W} \leq 0$) to hold when \eqref{eq:u_cond} and \eqref{eq:v_cond} hold. To satisfy these conditions, we can apply $\mathcal{S}$-procedure \cite[Sec. 2.6.3]{boyd1994linear}, which states that there should exist non-negative constants $\alpha$ and $\beta$ such that the following inequality holds:
		\begin{equation} 
			E_1 - \alpha F_1 - \beta G_1 \preceq 0.
		\end{equation}
		The condition above can be written as \eqref{eq:delta_locally_condition}. Note that in the derivation of \eqref{eq:delta_locally_condition}, we have imposed the condition on Lipschitz set condition in \eqref{eq:elipsoid_subset} and the norm constraint on parameters of nonlinear part of uncertainty model in \eqref{eq:theta_n_bound}. Therefore, the condition \eqref{eq:delta_locally_condition} together with \eqref{eq:elipsoid_subset}, and \eqref{eq:theta_n_bound} imply that the ellipsoid $\mathcal{E}_{inv}$ (as defined in \eqref{eq:model_set}) is an invariant set of the extended model \eqref{eq:model}.
		
		\section{ISS conditions for model \eqref{eq:model}}  \label{ap:iss_condtions}
		
		%Let us first introduce the following lemma, which is used to ensure ISS using an ISS Lyapunov function.
		%
		%\begin{lem}\emph{\textbf{(ISS Lyapunov Function~{\cite[Thm. 4.19]{khalil2002nonlinear}})}
		%		Consider the model \eqref{eq:model} and let $W(e)$ be a continuously differentiable function such that
		%		\begin{equation*}
			%			\alpha_{1}(\|e\|) \leq W(e) \leq \alpha_{2}(\|e\|),
			%		\end{equation*}
		%		\begin{equation*}
			%			\dot{W}(e) \leq-W_{3}(e), \quad \hspace{-1mm} \forall \hspace{1mm}\|e\| \geq \xi (\| \bar \omega_a (t) \|),
			%		\end{equation*}
		%		where $\alpha_{1}(\cdot)$ and $\alpha_{2}(\cdot)$ are class $\mathcal{K}_{\infty}$ functions, $\xi(\cdot)$ is a class $\mathcal{K}$ function, and $W_{3}$ is a continuous positive definite function. Then, the estimation error dynamics \eqref{eq:error_dynamics_combined} is ISS with ISS gain $ \mu(\|\bar \omega_a\|) = \alpha_{1}^{-1}(\alpha_{2}(\xi(\|\bar \omega_a\|)))$.}
	%	\label{lem: iss}
	%\end{lem}
	
	Let $W(x):={x}^\top P {x}$ be an ISS Lyapunov function candidate. Similar to the proof in Appendix \ref{ap:invariance_condtions}, given the globally Lipschitz property of the model nonlinearities in \eqref{eq:globally_lipschitz} and the imposed condition on parameters of nonlinear part of uncertainty model in \eqref{eq:theta_n_bound}, the time-derivative of the candidate ISS Lyapunov function, $\dot{W}$, can be written in the form of \eqref{eq:lyapanov_ap}. The rest of the proof follows the line of reasoning of the proof of Proposition~1 in \cite{ghanipoor2023robust}, which for brevity is omitted here. It is shown in \cite{ghanipoor2023robust} for a Lyapunov derivative inequality similar to \eqref{eq:lyapanov_ap} that, in order to conclude ISS property of a dynamics, $\Delta$ has to be negative definite. Note that in the derivation of $\Delta$ in \eqref{eq:lyapanov_ap}, we have imposed the condition on the parameters of nonlinear part of uncertainty model in \eqref{eq:theta_n_bound}. Therefore, the condition $\Delta \prec 0$ together with \eqref{eq:theta_n_bound} imply that the extended model \eqref{eq:model} is ISS with respect to input $u$.

			\section{Proof of Theorem \ref{theorem:learning_modified_cost_locally}}  \label{ap:thm1_proof}
			We show that the non-convex optimization problem in \eqref{eq:non_convex_optimization_loaclly_lip} can be convexified as in \eqref{eq:sdp_quadratic_cost_locally}. First, to make the problem tractable, we set the $S_{\eta_l} = I$ in extended system model \eqref{eq:model}. Then, we convexify the non-convex conditions in \eqref{eq:non_convex_optimization_loaclly_lip} as follows.  
			
			Consider the stability constraint \eqref{eq:delta_locally_condition}. By applying changes of variables as $S := P \Theta_l$, $R := P B_l$ and the Schur complement, the stability constraint in \eqref{eq:delta_locally_condition} is equivalent to \eqref{eq:delta_convex_locally}. Favorably, by the introduced change of variables, the stability constraint becomes convex (an LMI). However, since the cost $J$ in \eqref{eq:quadratic_cost} is a function of $\Theta_l = P^{-1} S$ and $B_l = P^{-1} R$, the cost is not convex in $S, R$ and $P$, after the changes of variables. Therefore, we have to convexify the cost to arrive at a convex optimization problem formulation. For scalar cost function $J$ in \eqref{eq:quadratic_cost}, we have the following equality:
			$$
			J = \sum_{i = 1}^{N} d_i^\top T^\top T d_i = \sum_{i = 1}^{N} tr(d_i^\top T^\top T d_i).
			$$
			Due to cyclic property of the trace operator, the above cost can be written as  
			$$
			J = tr(T^\top T \sum_{i = 1}^{N} d_i d_i^\top). 
			$$
			By the definition of the matrix $D$ in \eqref{eq:data_matrix} and the cyclic property of the trace, we have
			$$
			J = tr(T D T^\top).
			$$
			Now, we can write the epigraph form of the optimization problem in \eqref{eq:non_convex_optimization_loaclly_lip} as follows:
			\begin{mini}|s|[2]         	% mini! = minimize 
				{\substack{P,S, R, \Theta_n, W, \alpha, \gamma}}                             % optimization variable
				{tr(W)}   								% objective function and label
				{\label{eq:optimization_in_thm1_proof}}             								% label for optimizatio problem
				{}                           								% optimization result
				%	\addConstraint{}
				\addConstraint{\eqref{eq:delta_convex_locally},\eqref{eq:elipsoid_subset}, \eqref{eq:theta_n_bound}}{}   					  	% constraint 1
				\addConstraint{P}{\succ 0, \alpha, \gamma \geq 0}   					  	% constraint 3
				\addConstraint{tr(T D T^\top)}{\leq tr(W).}   					  	% constraint 3
			\end{mini}
			Due to monotonicity of trace, the last constraint in \eqref{eq:optimization_in_thm1_proof} can be transformed to the following constraint:
			\begin{equation*} 
				T D T^\top \preceq W.
			\end{equation*}
			Note that even with the application of the above transformation, the optimization remains equivalent to \eqref{eq:optimization_in_thm1_proof} \cite[p.8]{boyd1994linear}. By applying the Schur complement to the above inequality, we have 
			\begin{equation*} 
				\left[\begin{array}{cc}
					W & T\tilde{D}^\top \\
					\star & I
				\end{array}\right] \succeq 0.
			\end{equation*}
			Note that by construction, the data matrix $D$ is always symmetric and positive semi-definite. Therefore, its Cholesky decomposition (i.e., $D = \tilde{D}^\top \tilde{D}$) always exists. By applying the congruence transformation of $\text{diag}(P,I)$ to the above inequality, we obtain the following equivalent inequality:
			\begin{equation} \label{eq:GDG}
				\left[\begin{array}{cc}
					P W P &  \tilde{T} \tilde{D}^\top \\
					\star & I
				\end{array}\right] \succeq 0.
			\end{equation}
			Note that in the above inequality to introduce $\tilde{T} := P T = {\left[\begin{array}{cccc}
					S  & R & -P & P \Theta_n
				\end{array}\right]}$, the change of variable as $Z := P \Theta_n$ is applied.  Now, by substituting the lower bound $2\mu_2 P-\mu_2^2 W^{-1}$, with given positive scalar $\mu_2$, for $P W P$ and applying the Schur complement, the LMI in \eqref{eq:cost_related_constraint_locally} is obtained. 
			
			Consider another non-convex condition (upper bound on the parameters of nonlinear part of uncertainty model) as \eqref{eq:theta_n_bound}. Based on \cite[sec. 2.11]{caverly2019lmi}, the condition \eqref{eq:theta_n_bound} can be written as
			\begin{equation} \label{eq:theta_n_bound_matrix_form}
				\left[\begin{array}{cc}
					\bar l_{h_x} I &  l_{h_x} \Theta_n^\top \\
					\star & \bar l_{h_x} I
				\end{array}\right] \succeq 0.
			\end{equation}
			By applying the congruence transformation of $\text{diag}(I,P)$ to the above inequality, we obtain the following equivalent inequality
			\begin{equation*} 
				\left[\begin{array}{cc}
					\bar l_{h_x} I &  l_{h_x} Z^\top \\
					\star & \bar l_{h_x} P^2
				\end{array}\right] \succeq 0.
			\end{equation*}
			Now, by substituting the lower bound $2\mu_1 P-\mu_1^2 I$, with given positive scalar $\mu_1$, for $P^2$, the LMI in \eqref{eq:theta_n_bound_convex} is obtained.

			In conclusion, instead of the non-convex optimization problem in \eqref{eq:non_convex_optimization_loaclly_lip}, we provide an approximation of that problem in the form of the semi-definite program in \eqref{eq:sdp_quadratic_cost_locally}. We remark that the approximation arises from initially setting $S_{\eta_l} = I$ at the beginning of the proof. Additionally, we use the lower bound of $2\mu_2 P-\mu_2^2 W^{-1}$ for $P W P$ in the derivation of the LMI in \eqref{eq:cost_related_constraint_locally}, and the lower bound of $2\mu_1 P-\mu_1^2 I$ for $P^2$ in the derivation of the LMI in \eqref{eq:theta_n_bound_convex}.

			\section{Proof of Theorem \ref{theorem:learning_modified_constraint_locally}}  \label{ap:thm3_proof}
			Here, we follow the same line of reasoning as in the proof of Theorem \ref{theorem:learning_modified_cost_locally} by showing that the non-convex optimization problem in \eqref{eq:non_convex_optimization_loaclly_lip} can be convexified as in \eqref{eq:sdp_quadratic_constraint_locally}. First, we set $S_{\eta_l} = S_\eta$ in extended model \eqref{eq:model} (i.e., here we use the knowledge of uncertainty structure). Then, we convexify the non-convex conditions as follows. 
			
			Consider the constraint \eqref{eq:delta_locally_condition}. By applying the congruence transformation of $\text{diag}(P^{-1},I,I)$ to \eqref{eq:delta_locally_condition}, we obtain the following equivalent inequality:
			\begin{equation} \label{eq:Q_form_delta_condition_locally}
				\begin{aligned}
					\left[\begin{array}{ccc}
						Q \Delta Q +\beta Q  & B_u+S_{\eta} B_l & 0 \\
						\star & (l_{g_u} + \bar l_{h_u}) I - \alpha U & 0 \\
						\star & \star & \alpha - \beta
					\end{array}\right] \prec 0, 
				\end{aligned}
			\end{equation}
			where $Q := P^{-1}$. Using \eqref{eq:delta_globally}, the term $Q \Delta Q+\beta Q $ in the above inequality can be written as follows:
			\begin{equation} \label{eq:Q_form_delta_with_beta}
				\begin{aligned}
					Q \Delta Q+\beta Q & = A Q +Q A^\top+ Q V_\eta ^\top \Theta_l^\top S_{\eta}^\top + S_{\eta} \Theta_l V_\eta Q + (l_{g_x} +l_{g_u}) S_g S_g^\top   \\
					&+ (\bar l_{h_x}+ \bar l_{h_u}) S_{\eta_l} S_{\eta_l}^\top +l_{g_x} Q V_g^\top V_g Q + \bar l_{h_x} Q  V_\eta^\top V_\eta Q.
				\end{aligned}
			\end{equation}
			Using Young's inequality $\big(X^\top Y+Y^\top X \preceq \frac{1}{2}(X+\bar Z Y)^\top \bar Z^{-1} (X+\bar Z Y)$, with a symmetric positive definite matrix $\bar Z\big)$, we can find the following upper bound for \eqref{eq:Q_form_delta_with_beta}: 
			\begin{equation*} 
				\begin{aligned}
					Q \Delta Q+\beta Q &\leq A Q +Q A^\top +  \frac{1}{2}(V_\eta ^\top \Theta_l^\top S_\eta^\top +\bar Z Q)^\top \bar Z^{-1} (V_\eta ^\top \Theta_l^\top S_\eta^\top +\bar Z Q)  \\
					& + (l_{g_x} +l_{g_u}) S_g S_g^\top + (\bar l_{h_x} + \bar l_{h_u}) S_{\eta_l} S_{\eta_l}^\top +l_{g_x} Q V_g^\top V_g Q+ \bar l_{h_x} Q  V_\eta^\top V_\eta Q.
				\end{aligned}
			\end{equation*}
			
			Now, using the above upper bound for $Q \Delta Q+\beta Q$ and the Schur complement, the following sufficient condition can be found for the inequality in \eqref{eq:Q_form_delta_condition_locally}:
			\begin{equation*} 
				\left[\begin{array}{cccccc}
					N_{11}  & N_{12} & 0 &S_\eta \Theta_l V_\eta +Q \bar Z  &  N_{15} & N_{16}\\
					\star   & N_{22} & 0 & 0 & 0 & 0 \\
					\star   &  \star & N_{33} & 0 & 0 & 0\\
					\star  & \star  &  \star  &  -2 \bar Z & 0 & 0 \\
					\star  & \star  &  \star  & \star &  -I & 0 	\\
					\star  & \star  &  \star  &  \star &  \star & -I
				\end{array}\right] \prec 0
			\end{equation*}
			with $N_{11}, N_{12}, N_{15}, N_{16}, N_{22}$ and $N_{33}$ as defined in \eqref{eq:delta_lmi_components_constraint_app}. Given nonlinear term $Q\bar Z$ in the above inequality we select $\bar Z = \bar \gamma I$, with a positive scalar $\bar \gamma$. Therefore, the above inequality can be written as \eqref{eq:delta_convex_locally_constraint_app}. The non-convex condition (upper bound on the parameters of nonlinear part of uncertainty model) given in \eqref{eq:theta_n_bound} can be written as \eqref{eq:theta_n_bound_convex_constraint_app}. For the details see derivation of \eqref{eq:theta_n_bound_matrix_form} in proof of Theorem \ref{theorem:learning_modified_cost_locally}.
			
			The Lipschitz set condition \eqref{eq:elipsoid_subset}, in terms of $Q = P^{-1}$ can be written as 
			\begin{equation*} 
				\left[\begin{array}{cc}
					\gamma   F -  Q^{-1} & 0 \\
					* & - \gamma +1
				\end{array}\right] \preceq 0.
			\end{equation*}
			
			Now, by substituting the upper bound $\gamma F -  2 \mu_3 I + \mu_3^2 Q$, with given positive scalar $\mu_3$, for $\gamma F -  Q^{-1}$, the LMI in \eqref{eq:elipsoild_subset_Q} is obtained. Furthermore, the quadratic cost can be treated as constraint \eqref{eq:cost_constraint_modification} by writing the epigraph form of the optimization problem (see \eqref{eq:optimization_in_thm1_proof} for the epigraph form in proof of Theorem \ref{theorem:learning_modified_cost_locally}). Thus, instead of the non-convex optimization problem in \eqref{eq:non_convex_optimization_loaclly_lip}, we provide the approximate semi-definite program in \eqref{eq:sdp_quadratic_constraint_locally}.  We remark that the approximation arises from using a sufficient condition (Young's inequality together with selection of $\bar Z = \bar \gamma I$) in the derivation of the LMI in \eqref{eq:delta_convex_locally_constraint_app}. Additionally, we use the upper bound $\gamma F -  2 \mu_3 I + \mu_3^2 Q$ for $\gamma F -  Q^{-1}$ in the derivation of the LMI in \eqref{eq:elipsoild_subset_Q}.
			
			\section{Roll Plane System Description}  \label{ap:rpm_app}
			The equations of motion of the roll plane system are given in \cite[app. B.2]{kessels2023real}. Here, we provide the state space representation. By defining the state vector $x_s = [x_{s_1}, x_{s_2},\ldots,x_{s_7}, x_{s_8}]^\top := [q_1, q_2, q_3, q_4,\dot{q}_1,\ldots, \dot{q}_4]^\top$, where $q_i$ and $\dot{q}_i$ are the displacements (as shown in Figure \ref{fig:rpm_schematic}) and the corresponding velocities, the system dynamics can be described in the form of \eqref{eq:sys}, where we have
			\begin{equation*} 
				\begin{aligned}
					A &=\left[\begin{array}{cc}
						0    & I \\
						-\tilde{M}^{-1}\tilde{K} & -\tilde{M}^{-1}\tilde{C} \end{array}\right], \quad 
					B_u =\left[\begin{array}{c}
						0 \\
						\tilde{M}^{-1} \tilde{K}_u
					\end{array}\right], \quad
					S_g = S_\eta =\left[\begin{array}{c}
						0 \\
						- \tilde{M}^{-1} \tilde{S}
					\end{array}\right], \\
					g(\cdot) &= 0.2 \tilde{c}_{1,n} tanh(V_g x_s) \quad
					D_\nu = I, \quad B_\omega= 0, \quad
					\eta(\cdot) = \delta \tilde{k}_1 V_{\eta} x_s + \tilde{k}_{1_n} (V_{\eta} x_s)^{\circ 3}, \\
					C&=\left[\begin{array}{cc}
						\tilde{C} & 0 \\
						0 & \tilde{C} 
					\end{array}\right], \quad 
					\tilde{C}=\left[\begin{array}{cccc}
						1 & 0 & -1 & 0 \\
						0 & 1 & 0 & -1 \\
					\end{array}\right] \quad
					V_g =\left[\begin{array}{cc}
						0 & 10 \tilde{C} 
					\end{array}\right], \quad
					V_\eta =\left[\begin{array}{cc}
						\tilde{C}  & 0 
					\end{array}\right], 
				\end{aligned}
			\end{equation*}
			\begin{equation*} 
				\begin{aligned}
					\tilde{M} &= \left[\begin{array}{cccc}
						\frac{\tilde{m}}{2} & \frac{\tilde{m}}{2} & 0 & 0 \\
						\frac{-\tilde{I}}{\tilde{L}} & \frac{\tilde{I}}{\tilde{L}} & 0 & 0 \\
						0 & 0 & \tilde{m}_{t_1} & 0 \\
						0 & 0 & 0 & \tilde{m}_{t_2}
					\end{array}\right], \quad
					\tilde{K} = \left[\begin{array}{cccc}
						\tilde{k}_1 & \tilde{k}_2 & -\tilde{k}_1 & -\tilde{k}_2 \\
						\frac{-\tilde{L}}{2}\tilde{k}_1 & \frac{\tilde{L}}{2}\tilde{k}_2 & \frac{\tilde{L}}{2}\tilde{k}_1 & \frac{-\tilde{L}}{2}\tilde{k}_2 \\
						-\tilde{k}_1 & 0 & \tilde{k}_1+\tilde{k}_{t_1} & 0 \\
						0 & -\tilde{k}_2 & 0 & \tilde{k}_2+\tilde{k}_{t_2}
					\end{array}\right], \\
					\tilde{C} &= \left[\begin{array}{cccc}
						\tilde{c}_1 & \tilde{c}_2 & -\tilde{c}_1 & -\tilde{c}_2 \\
						\frac{-\tilde{L}}{2}\tilde{c}_1 & \frac{\tilde{L}}{2}\tilde{c}_2 & \frac{\tilde{L}}{2}\tilde{c}_1 & \frac{-\tilde{L}}{2}\tilde{c}_2 \\
						-\tilde{c}_1 & 0 & \tilde{c}_1 & 0 \\
						0 & -\tilde{c}_2 & 0 & \tilde{c}_2
					\end{array}\right], \quad
					\tilde{K}_u =\left[\begin{array}{cc}
						0 & 0\\
						0 & 0 \\
						\tilde{k}_{t_1} & 0 \\
						0 & \tilde{k}_{t_2}  \\
					\end{array}\right], \quad
					\tilde{S}=\left[\begin{array}{cc}
						1 & 1\\
						\frac{-\tilde{L}}{2} & \frac{\tilde{L}}{2} \\
						-1 & 0 \\
						0 & -1  \\
					\end{array}\right].
				\end{aligned}
			\end{equation*}
			
			The parameter values are listed in Table \ref{tab:rpm_parameters}.
			
			\begin{table}[b!] 
				\caption{Parameter values of roll plane system.}
				\centering
				\begin{tabular}{ccc}
					\hline
					Parameter                                                           & Value  & Unit   \\ \hline
					$\tilde{m}$ &               $580$                                                                      &   $\mathrm{~kg}$                                                                                                                      \\ 
					$\tilde{m}_{t_1}, \tilde{m}_{t_2}$ &                          $36.26$                                                            &   $\mathrm{~kg}$                                                                                                                                                \\ 
					$\tilde{I}$ &                          $63.3316$                                                            &   $\mathrm{~kg} \cdot \mathrm{m}^2$                                                                                                                                               \\
					$\tilde{L}$ &                          $1.524$                                                            &   $\mathrm{~m}$                                                                                                                                                \\ 
					$\tilde{c}_1, \tilde{c}_2$ &                          $710.70$                                                            &   $\frac{\mathrm{N} \cdot \mathrm{s}}{\mathrm{m}}$                                                                                                                                                \\ 
					$\tilde{c}_{1,n}, \tilde{c}_{2,n}$ &                  $0.71$                                                            &   $\frac{\mathrm{N} \cdot \mathrm{s}}{\mathrm{m}}$                                                                                                                                                \\ 
					$\tilde{k}_1, \tilde{k}_2$ &                          $19357.2$                                                            &   $\frac{\mathrm{N}}{\mathrm{m}}$    \\
					$\tilde{k}_{t_1}, \tilde{k}_{t_2}$ &                          $96319.76$                                                            &   $\frac{\mathrm{N}}{\mathrm{m}}$                                                                                                                                        \\
					$\tilde{k}_{1_n}, \tilde{k}_{2_n}$ &                          $15000$                                                            &   $\frac{\mathrm{N}}{\mathrm{m}^3}$                                                                                                                                         \\ 
					$\delta \tilde{k}_1, \delta \tilde{k}_2$ &                          $5807.2$                                                            &   $\frac{\mathrm{N}}{\mathrm{m}}$                                                                                                                                                \\ \hline
				\end{tabular} \label{tab:rpm_parameters}
			\end{table}
			
			\begin{figure}[htbp]
    	\begin{minipage}[t]{0.48\textwidth}
				\centering
				\smallskip
				\includegraphics[width=1\linewidth,keepaspectratio]{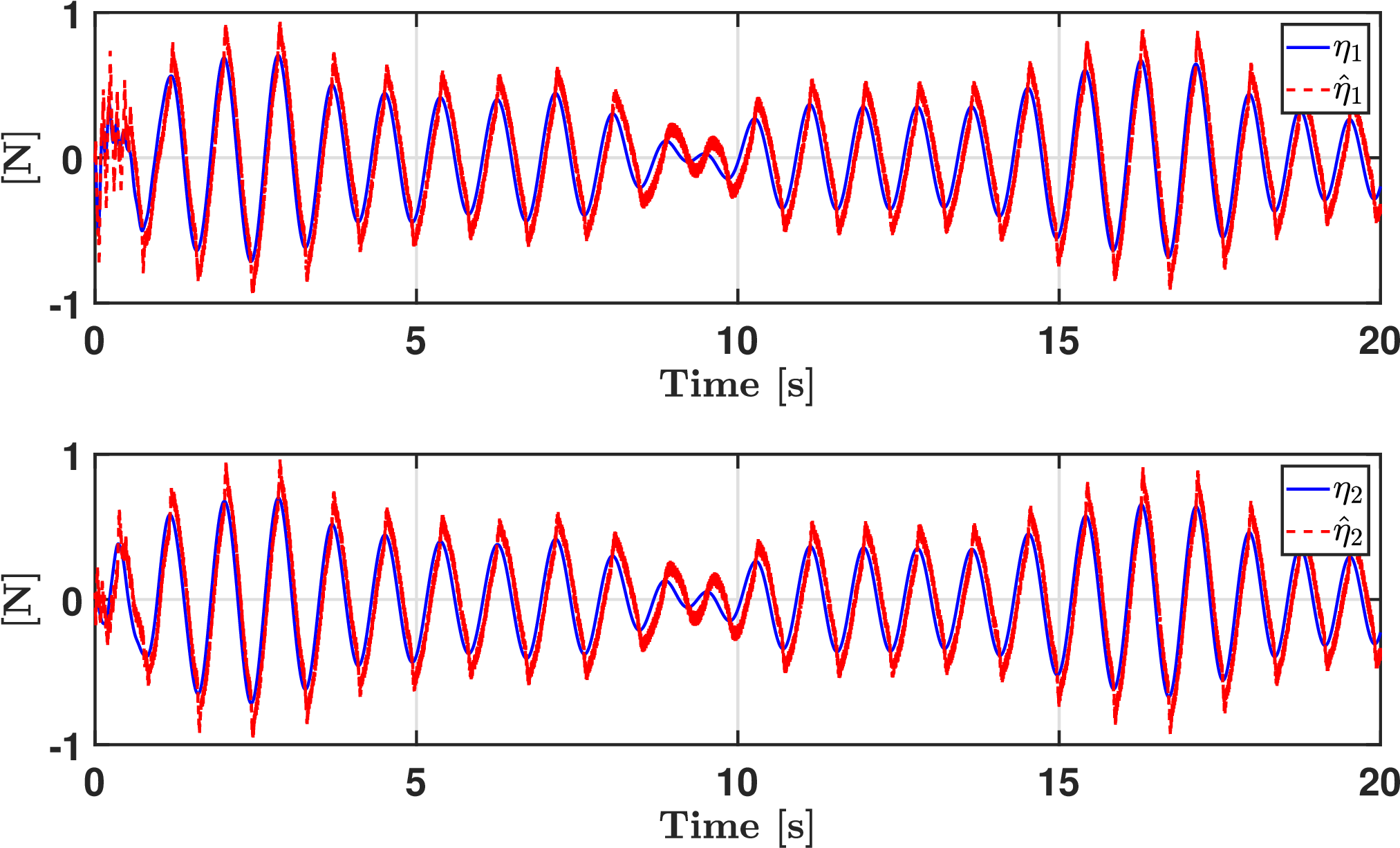}
				\caption{Uncertainty entries and their estimates  for a randomly selected test data.}
				\label{fig:uncertainty}
			\end{minipage}
			\hfill
    \begin{minipage}[t]{0.48\textwidth}
				\centering
				\smallskip
				\includegraphics[width=1\linewidth,keepaspectratio]{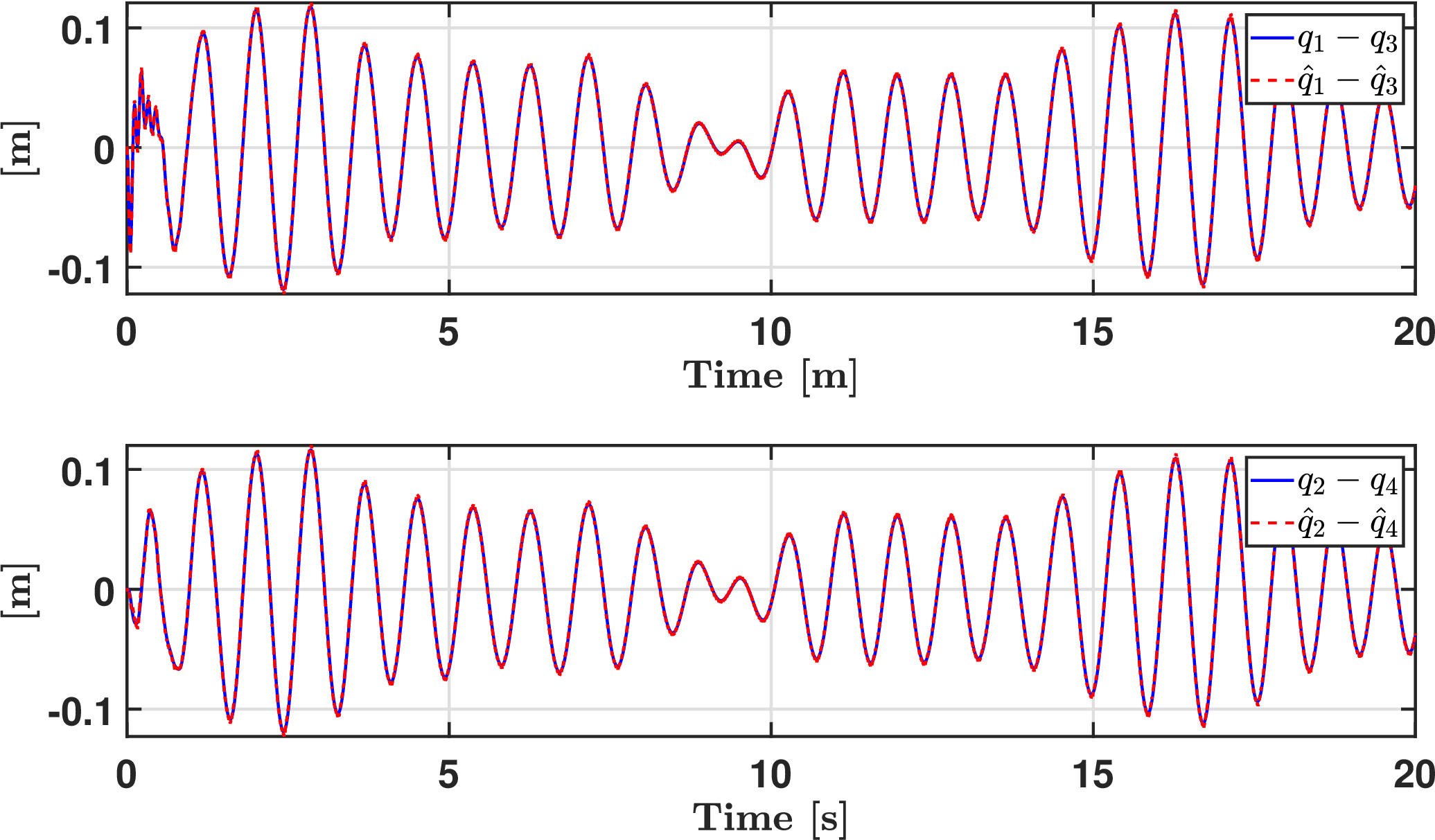}
				\caption{Deriving terms of the uncertainty $\eta(\cdot)$, i.e., $V_\eta x_s$ and their estimates, i.e., $V_\eta \hat{x}_s$, for a randomly selected test data.}
				\label{fig:state}
			\end{minipage}
		\end{figure}

			\section{Results for Uncertainty and State Estimation}  \label{ap:estimation_result}
			
			For one test data set in the considered case study (i.e., the roll plane system), the entries of uncertainty $\eta$ in \eqref{eq:sys} and their estimates are depicted in Figure \ref{fig:uncertainty}. Moreover, Figure \ref{fig:state} illustrates the deriving terms of the uncertainty $\eta(\cdot)$, i.e., $V_\eta x_s$ and their estimations, i.e., $V_\eta \hat{x}_s$. As demonstrated, the proposed uncertainty and state estimation method yield highly accurate results.

			\section*{Acknowledgements}                               % Place acknowledgements
				This publication is part of the project Digital Twin project
				4.3 with project number P18-03 of the research programme
				Perspectief which is (mainly) financed by the Dutch Research
				Council (NWO).
			
			\bibliographystyle{unsrt}        % Include this if you use bibtex 
			\bibliography{ref}           % and a bib file to produce the 

			%	\appendix
			%	\section{Appendices}
			%	\subsection{Lyapanov Function Proof} \label{ap: lyapanov}
			%	The inequality in \eqref{eq: lyapanov} can be derived as follows:

		\end{document}